\documentclass[aps,pra,10pt,showpacs
]{revtex4}
\usepackage{graphicx}
\usepackage{amsmath,amsthm}
\usepackage[active]{srcltx} 
\begin{document}
\title{Low-Lying Excitation Modes of Trapped Dipolar Fermi Gases: \\ From Collisionless to Hydrodynamic Regime}
\author{Falk W\"{a}chtler}
\affiliation{Institut f\"ur Theoretische Physik, Leibniz Universit\"at Hannover, Germany}
\author{Aristeu R. P. Lima}
\email{aristeu@unilab.edu.br}

\affiliation{Universidade da Integra\c{c}\~{a}o Internacional da Lusofonia Afro-Brasileira, Campus dos Palmares, 
62785-000 Acarape-Cear\'{a}-Brazil}
\author{Axel Pelster}
\email{axel.pelster@physik.uni-kl.de}
\affiliation{Fachbereich Physik und Forschungszentrum OPTIMAS, Technische Universit\"at Kaiserslautern, Germany}
\affiliation{Hanse-Wissenschaftskolleg, Lehmkuhlenbusch 4, 27733 Delmenhorst, Germany}

\begin{abstract}
By means of the Boltzmann-Vlasov kinetic equation we investigate dynamical properties of a trapped, one-component Fermi gas at zero temperature, featuring the anisotropic and long-range dipole-dipole interaction. To this end, we determine an approximate solution by rescaling both space and momentum variables of the equilibrium distribution, thereby obtaining coupled ordinary differential equations for the corresponding scaling parameters. Based on previous results on how the Fermi sphere is deformed in the hydrodynamic regime of a dipolar Fermi gas, we are able to implement the relaxation-time approximation for the collision integral. Then, we proceed by linearizing the equations of motion around the equilibrium in order to study both the frequencies and the damping of the low-lying excitation modes all the way from the collisionless to the hydrodynamic regime. Our theoretical results are expected to be relevant for understanding current experiments with trapped dipolar Fermi gases.
\end{abstract}
\maketitle
\section{Introduction}
The experimental achievement of Bose-Einstein condensation (BEC) of atomic chromium \cite{PhysRevLett.94.160401}, which has a magnetic moment of $m=6\mu_{\rm B}$, where $\mu_{\rm B}$ stands for the Bohr magneton, and the subsequent detection of the long-range and anisotropic dipole-dipole interaction (DDI) in that system \cite{stuhler_j_2005} paved the way for a systematic investigation of dipolar quantum systems. In chromium BECs, the DDI is usually of secondary importance, as its relative strength in comparison with the short-range and isotropic contact interaction is only of about $15\%$. However, by using a Feshbach resonance, one can tune the contact interaction in order to improve the relative importance of the DDI \cite{PhysRevLett.94.183201}. Thus, in retrospect, the work with chromium has turned out to be highly valuable as many specific achievements have led to 
a better understanding of bosonic dipolar systems \cite{baranov-review,citeulike:4464283,1367-2630-11-5-055009,doi:10.1021/cr2003568}. Among them, 
one could emphasize the observation of a d-wave Bose-nova explosion pattern \cite{d-wave-pfau}, the strong dipolar character of the time-of-flight analysis \cite{strong-pfau}, and the detection of the influence of the DDI in the low-lying excitations \cite{PhysRevLett.105.040404,gorceix_sound}.

Since then, a few major experimental successes were obtained, which might lead to even stronger dipolar quantum systems, consisting of both atomic and molecular systems. Particularly promising candidates are polar molecules such as KRb \cite{K.-K.Ni10102008,citeulike:6565167,efficient_transfer}, LiCs \cite{weidemueller1,weidemueller2} and, most recently obtained, NaK \cite{park}, which have a strong electric dipole moment.Their DDI may be up to ten thousand times stronger than in usual atomic systems \cite{arXiv:0811.4618} with the additional property of tunability of strength, sign, and direction \cite{silke_tune}, thus providing an ideal testing ground for strong dipolar systems. Moreover, atomic systems such as dysprosium, the most magnetic atom with a magnetic moment of $m=10\mu_{\rm B}$, and erbium, which has $m=7\mu_{\rm B}$,  are currently under intense investigation. Indeed, the bosonic dysprosium isotope $^{164}$Dy was Bose condensed \cite{PhysRevLett.107.190401}, while its fermionic isotope $^{161}$Dy was brought to quantum degeneracy \cite{PhysRevLett.108.215301} and shown to display a fermionic suppression of the inelastic dipolar scattering \cite{PhysRevLett.114.023201}. On top of that, Bose-Einstein condensation of $^{168}$Er was achieved \cite{PhysRevLett.108.210401} in which the long-sought roton mode \cite{santos_roton} could recently be observed \cite{roton}. Furthermore, also a quantum degenerate dipolar Fermi gas of $^{167}$Er atoms was created \cite{Ferlaino} which was used for the remarkable experimental achievement of demonstrating the deformation of the Fermi sphere in a dipolar Fermi gas \cite{ferlaino_deform}.  In this context, fascinating possibilities are brought about by the recent production of weakly bound molecular states in Er$_{2}$ molecules, which have very large dipole moments and whose orientation can be changed \cite{er_molecule}. Naturally, such experimental developments have triggered much theoretical interest such as in dipolar Bose-Fermi mixtures \cite{adhikari}, two-component dipolar Fermi gases \cite{bienias}, exact ground-state \cite{amit} and quasiparticle properties \cite{kopietz} as well as few fermion quench dynamics in one-dimension optical lattices \cite{grass}.

The combination of theoretical and experimental interest in highly magnetic atoms has also led to a major development in dipolar quantum gases which is the demonstration that a strongly dipolar Bose gas can exhibit phenomena such as the Rosensweig instability \cite{pfau_rosensweig} by means of the formation of quantum droplets in its ground state instead of a mere condensate \cite{pfau_droplets} and that these might even unite into a single large droplet \cite{russel1,ferlaino_big_droplet}. Indeed, by taking into account the influence of quantum fluctuations of the ground-state energy \cite{quantumfluctuations,beyondmeanfield} within the non-linear non-local Gross-Pitaevskii equation can explain both the small droplets in dysprosium \cite{falk_droplets} and the large ones in erbium \cite{ferlaino_big_droplet} without any fitting parameter. Related results were also obtained by using quantum Monte Carlo simulations regarding droplets \cite{mazzanti} but also similar types of structures such as filaments \cite{tommaso}. Recent experimental results even suggested that tilting the orientation of the dipoles induces a phase transition from a BEC to a metastable state of many tilted droplets \cite{pfau_striped}. Moreover, further theoretical investigation led to the nontrivial fact that these droplets are self-bound, i.e., remain localized when released in free space \cite{russel}.

Concerning dipolar systems, one feature, which has received considerable attention, is the possibility of magnetostriction in momentum space. Indeed, the recent experimental observation of this effect, realized with atomic erbium samples \cite{ferlaino_deform}, provides a long awaited confirmation of theoretical predictions and, at the same time, paves the way for a plethora of other related physical phenomena. While being first found theoretically in the fermionic case \cite{miyakawa_t_2008}, it was intensively investigated in different contexts and also in the 
bosonic case \cite{PhysRevA.86.023605}. For fermionic dipolar gases, other interesting possibilities were found such as 
supersolid \cite{PhysRevA.83.053629}, ferronematic \cite{fregoso:205301}, and Berezinskii-Kosterlitz-Thouless phases \cite{bruun:245301} and in addition also their Fermi liquid properties were studied \cite{PhysRevA.81.023602}.

Much theoretical work has also been devoted to the normal phase of a dipolar Fermi gas. Its equilibrium properties in the presence of a harmonic trap was 
considered at both zero \cite{rzazewski,miyakawa_t_2008,AristeuRapCom,Aristeu} and finite temperature \cite{endo,PhysRevA.81.033617}. Also the 
dynamical properties of a trapped dipolar Fermi gas were studied at both zero \cite{1367-2630-11-5-055017,AristeuRapCom,Aristeu} and finite
temperature \cite{PhysRevA.83.053628}.

Dynamical properties such as the low-lying excitations represent an important diagnostic tool for ultracold systems. Moreover, they can be 
measured with high accuracy, so as to provide reliable physical information. In the case of non-dipolar unitary Fermi gases, for example, such 
measurements were used to discard predictions of the mean-field Bardeen-Cooper-Schrieffer theory (BCS) in favor of the predictions 
of quantum Monte Carlo simulations \cite{PhysRevLett.95.030404} along the BCS-BEC crossover \cite{PhysRevLett.98.040401}. A more recent example is 
that of the experimental support for the lower bound to the viscosity of an unitary gas \cite{Cao07012011}, which is conjectured to be 
universal \cite{PhysRevLett.94.111601}. The aforementioned detection of the DDI through observing the hydrodynamic modes of a chromium BEC should also 
be recalled in this context \cite{stuhler_j_2005}.

In the first studies, the investigations of the excitations of dipolar Fermi gases were concentrated on either the collisionless (CL) regime 
\cite{1367-2630-11-5-055017,PhysRevA.83.053628}, where collisions can be neglected, or in the hydrodynamic (HD) regime \cite{AristeuRapCom,Aristeu},  
where collisions occur so often that local equilibrium can be assumed. Recently, also the radial quadrupole mode was studied in detail in 
both the CL and the HD regime \cite{PhysRevA.85.033639}. Moreover, numerical studies focusing on a two-dimensional system in the HD regime were performed \cite{zaremba}. The next natural step is the investigation of what happens, when the system undergoes a 
crossover from one regime to the other. Along these lines, there was recently a thorough investigation of quasi-two-dimensional dipolar Fermi gases \cite{demler}. 
Indeed, by considering a linearized scaling ansatz as well as numerical results, the first eight moments of the collisional Boltzmann-Vlasov kinetic equation 
were analyzed. The resulting graph of the collision rate against temperature was shown to exhibit an unexpected plateau, a unique 
characteristic of dipolar systems, and also the low-lying modes in this quasi two-dimensional case were considered \cite{demler}. It is important to 
remark that, additionally, 
the effect of quantum correlations was considered in this system. Building on a previous study, where quantum Monte Carlo methods 
were used to investigate the Fermi liquid as well as the crystal phases in strictly two-dimensional systems \cite{matveeva_prl}, a mapping scheme could 
be constructed, that allows for investigating  correlations in quasi two-dimensional, spherically trapped dipolar Fermi gases \cite{demler2}.

In the present paper, we focus on a three-dimensional dipolar Fermi gas at zero temperature and investigate the transition of both the frequencies 
and the damping of the low-lying modes from the CL to the HD regime by applying the relaxation-time approximation. To this end the paper 
is organized as follows. In Section \ref{BoVlEq} we solve the Boltzmann-Vlasov (BV) kinetic equation 
for a harmonically trapped dipolar Fermi gas by rescaling the equilibrium distribution and
obtain ordinary differential equations for the respective scaling parameters. Afterwards, we specialize them in Section \ref{equi} at zero temperature to 
the relaxation-time approximation for the collision integral and to a concrete equilibrium distribution.
A subsequent linearization of the equations of motion for the respective scaling parameters
in Section \ref{linear}
allows to determine both the frequency and the damping of the low-lying excitation modes. In particular, we investigate how
the properties of the monopole mode, the three-dimensional quadrupole mode, and the radial quadrupole mode change by varying the relaxation time
from the CL to the HD regime. The conclusion in Section \ref{CON} summarizes our findings and indicates possible future investigations
along similar lines. Furthermore, the
appendix presents a self-contained computation of the respective relevant energy integrals, as well as technical details concerning the linearization of the equations of motion.

\section{Boltzmann-Vlasov Equation}\label{BoVlEq}
We start with describing the dynamic properties of a trapped dipolar Fermi gas by means of the
Wigner function $\nu({\bf x},{\bf q},t)$, which represents a
semiclassical distribution function in phase space spanned by coordinate ${\bf x}$ and wave vector ${\bf q}$ \cite{Wigner1,Wigner2}.
It allows to determine both the particle density through $n({\bf x},t)=\int d^3q \,\nu({\bf x},{\bf q},t)/(2\pi)^3$
and the wave vector distribution by $n({\bf q},t)=\int d^3x\,\nu({\bf x},{\bf q},t)/(2\pi)^3$ as well as the expectation value of any
observable according to $\langle O \rangle= \int d^3x \int d^3q\,O({\bf x},{\bf q}) \nu({\bf x},{\bf q},t)/(2\pi)^3$. 
The time evolution of this Wigner function is 
determined by the Boltzmann-Vlasov kinetic equation 
\begin{equation}
  \frac{\partial \nu}{\partial t}+\left\{\frac{\hbar {\bf q}}{M}+\frac{1}{\hbar} \frac{\partial\left[U({\bf x})
+U_{\rm mf}({\bf x},{\bf q},t)\right]}{\partial 
{\bf q}}\right\} \frac{\partial \nu}{\partial {\bf x}}-\frac{1}{\hbar}\frac{\partial\left[U({\bf x})+U_{\rm mf}({\bf x},{\bf q},t)\right]}{\partial {\bf x}} 
\frac{\partial \nu}{\partial {\bf q}}=I_{\rm coll}[\nu]({\bf x},{\bf q},t) \, , \label{bve}
\end{equation}
where $U({\bf x})=M\sum_i \omega_i^2 x_i^2/2$ is a general harmonic trapping potential for a particle with mass $M$ and $\omega_i$ denotes the trap frequency in the $\imath$-direction. 
The mean-field potential 
\begin{equation} 
U_{\rm mf}({\bf x},{\bf q},t)=\int d^3x'n({\bf x'},t)V_{\rm d}({\bf x}-{\bf x'}) - \int \frac{d^3q'}{(2\pi)^3}\nu({\bf x},{\bf q'},t)
\tilde{V}_{\rm d}({\bf q}-{\bf q'})
\label{umf}
\end{equation}
contains in the first and second term the Hartree and Fock contributions, respectively, where
$V_{\rm d}({\bf x})$ represents the dipole-dipole potential and $\tilde{V}_{\rm d}({\bf q})$ its Fourier transform. We consider a system of dipolar 
fermions with the point dipoles aligned along the $z$-direction so that $V_{\rm d}({\bf x})$ reads
\begin{equation}
V_{\rm d}({\bf x})=\frac{C_{\rm dd}}{4\pi \left|{\bf x}\right|^3}\left(1-3 \cos^2 \vartheta \right)\,, 
\label{dipolepotential}
\end{equation}
with $\vartheta$ being the angle between the direction of the polarization of the dipoles and their relative position. The dipole-dipole interaction 
strength is 
characterized for magnetic atoms by $C_{\rm dd}=\mu_0m^2$, with $\mu_0$ being the magnetic permeability in vacuum and $m$ denoting the magnetic dipole 
moment. In the case 
of heteronuclear molecules with electric moment $d$, the interaction strength is given by $C_{\rm dd}=d^2/{\rm \epsilon_0}$, with the vacuum dielectric 
constant  
$\epsilon_0$. Note that
the Fourier transform of the dipole-dipole interaction potential (\ref{dipolepotential}) is given by \cite{goral}
\begin{equation}
\tilde{V}_{\rm d}({\bf k})=\int d^3x V_{\rm d}({\bf x}){\rm e}^{i{\bf k}\cdot
 {\bf x}}=\frac{C_{\rm dd}}{3}\left( \frac{3 k_z^2}{{\bf k}^2}-1\right)\,.
\label{FTDDP}
\end{equation}
The collision integral $I_{\rm coll}[\nu]$ includes the dissipative effects by means of a nonlinear functional of the distribution function and is 
of second order of the dipole-dipole potential $V_{\rm d}({\bf x})$. Its concrete 
form and derivation for a general two-body interaction potential can be found, for instance, in Ref.~\cite{KadanoffBaym}.

In order to find an approximate solution of the BV equation in the vicinity of 
equilibrium we use the scaling method from Ref.~\cite{String}. To this end, we assume that 
the distribution function $\nu({\bf x},{\bf p},t)$ can be obtained from rescaling the equilibrium distribution function $\nu^0({\bf r},{\bf k})$, 
which satisfies 
\begin{equation}
\left\{\frac{\hbar {\bf q}}{M}+\frac{1}{\hbar} \frac{\partial\left[U({\bf x})+U_{\rm mf}({\bf x},{\bf q})\right]}{\partial {\bf q}}\right\} 
\frac{\partial \nu^0}{\partial {\bf x}}-\frac{1}{\hbar}\frac{\partial\left[U({\bf x})+U_{\rm mf}({\bf x},{\bf q})\right]}{\partial {\bf x}} 
\frac{\partial \nu^0}{\partial {\bf q}}=0\,,
\label{BVEE}
\end{equation}
according to
\begin{equation}
\nu({\bf x},{\bf q},t) = \Gamma \nu^0\left({\bf r}(t),{\bf k}(t)\right)\,.
\label{scalingWigner}
\end{equation}
Thereby we have introduced the scaling parameters $b_{\imath}$ and $\Theta_{\imath}$ via
\begin{eqnarray}
r_i &=&{\displaystyle \frac{x_i}{b_{\imath}(t)}}\,, 
\label{b}\\
k_i &=&{\displaystyle \frac{1}{\sqrt{\Theta_{\imath} (t)}}\left[ q_i-\frac{M\dot{b}_{\imath}(t) x_i}{\hbar b_{\imath}(t)}\right]}\,, 
\label{scalingwignerk}
\end{eqnarray}
where the second term in Eq.~(\ref{scalingwignerk}) describes a transformation of the ${\bf q}$-vector resulting in a vanishing 
local velocity field \cite{CastinDum}. As the scaling parameters 
$b_{\imath}$ and $\Theta_{\imath}$ denote the time-dependent deviation from equilibrium, their equilibrium values are given by $b_{\imath}^0=\Theta_{\imath}^0=1$. 
Furthermore, the term 
\begin{equation}
\Gamma={\displaystyle \frac{1}{\prod_j b_j \sqrt{\Theta_{\jmath}}}}
\end{equation}
ensures the normalization of the distribution function $\nu$. 

Ordinary differential equations for the scaling parameters $b_{\imath}$ and $\Theta_{\imath}$ can be obtained by taking moments of the BV equation, i.e.~by integrating it 
with a 
prefactor over the whole phase space. Using the prefactor $r_ik_i$, i.e.~performing the operation $\int d^3rd^3k/(2\pi)^3r_ik_i \,\times$ (\ref{bve}), 
leads to the following coupled ordinary differential equations for the spatial scaling parameters $b_{\imath}$  
\begin{align}
&\ddot{b}_{\imath}+  \omega_i^2 b_{\imath}-\frac{\hbar^2 \left\langle k_i^2 \right\rangle^0 \Theta_{\imath}}{M^2 b_{\imath} \left\langle r_i^2 \right\rangle^0}
+\frac{1}{2M b_{\imath} \left\langle r_i^2 
\right\rangle^0}\left[\int \frac{d^3k}{(2\pi)^3}\tilde{W}_i(b,{\bf k}) \tilde{n}^0({\bf k}) \tilde{n}^0(-{\bf k})\right. \nonumber \\ 
&\hspace{1cm}\left. -\frac{1}{\prod_j b_j}\int \frac{d^3rd^3kd^3k'}{(2\pi)^6}\nu^0({\bf r},{\bf k})\nu^0({\bf r},{\bf k'}) \tilde{W}_i(\Theta,{\bf k}
-{\bf k'})\right]=0\,,  
\label{diffglb} 
\end{align}
where $\left\langle \bullet \right\rangle^0=\int d^3rd^3k\,\bullet \nu^0({\bf r},{\bf k})/(2\pi)^3$ denotes the phase-space average in equilibrium 
and $\tilde{n}^0$ is the 
Fourier transform of the spatial density in equilibrium $n^0({\bf r})=\int d^3q\,\nu^0({\bf r},{\bf q})/(2\pi)^3$. 
Furthermore $\tilde{W}_i(b,{\bf k})$ represents an abbreviation for
\begin{equation}
\tilde{W}_i(b,{\bf k})=F\left[r_i \frac{\partial V_{\rm d}(b,{\bf r})}{\partial r_i}\right]\,, 
\label{fouriertransformedhartreepotential}
\end{equation}
where $V_{\rm d}(b,{\bf r})$ stands for the rescaled dipole-dipole potential and $F[\bullet]$ the Fourier transform. The other abbreviation function
$\tilde{W}_i(\Theta,{\bf k})$ is defined by
\begin{equation} 
\tilde{W}_i(\Theta,{\bf k}-{\bf k'})=F\left[ r_i\frac{\partial V_{\rm d}({\bf r})}{\partial r_i}\right]\left(\Theta_x^{\frac{1}{2}}(k_x-k_x'),
\Theta_y^{\frac{1}{2}}(k_y-k_y'),\Theta_z^{\frac{1}{2}}(k_z-k_z')\right)\,. 
\label{fouriertransformedfockpotential}
\end{equation}
We observe that Eq.~(\ref{diffglb}) has a Newtonian 
form stemming from trapping, kinetic, and Hartree-Fock mean-field energy terms. Note that the contribution of the collision integral in (\ref{diffglb}) 
vanishes, 
because the quantity $r_ik_i$ is conserved under collisions \cite{String}.
The effect of collisions is only contained in the differential equations for the 
momentum scaling parameters $\Theta_{\imath}$, which can be obtained by taking moments of (\ref{bve}) with the prefactor $k_i^2$, leading to
\begin{equation}
\frac{\dot{\Theta}_i}{\Theta_{\imath}}+2\frac{\dot{b}_{\imath}}{b_{\imath}}=\frac{1}{\Gamma 
\left\langle k_i^2 \right\rangle^0}\int \frac{d^3rd^3k}{(2\pi)^3}k_i^2I_{\rm coll}[\nu]\,. 
\label{diffglT}
\end{equation}
We remark that taking moments of the BV equation weighted with a prefactor of the form $r_ir_j$ does not provide new constraints between 
the scaling parameters $b_{\imath}$ and $\Theta_{\imath}$ 
\cite{Dusling}. 
\section{Equilibrium}\label{equi}
Before solving the coupled set of
Eqs.~(\ref{diffglb}) and (\ref{diffglT}) we have to simplify them by explicitly evaluating the respective integrals. 
This can be done analytically by determining 
the equilibrium distribution function $\nu^0({\bf r},{\bf k})$ in a self-consistent way within a variational ansatz. In the low-temperature regime 
it is appropriate to 
choose an ansatz which resembles the Fermi-Dirac distribution of a non-interacting Fermi gas at zero temperature:
\begin{equation}
\nu^0({\bf r},{\bf k})=\boldsymbol\Theta \left(1-\sum_{\jmath}\frac{r_j^2}{R_{\jmath}^2}-\sum_{\jmath} \frac{k_j^2}{K_j^2}\right)\,
\label{ansatzwigner}
\end{equation}
with $\boldsymbol\Theta(x)$ denoting the Heaviside unit step function. Here the variational parameters $R_{\imath}$ and $K_{\imath}$ represent the Thomas-Fermi radii and momenta, respectively, which describe the extension of the 
equilibrium Fermi surface 
in both coordinate 
and momentum space. The dipole-dipole interaction stretches both the particle density \cite{rzazewski} and the momentum distribution 
\cite{miyakawa_t_2008} in the direction of the polarization, which is taken into account by a
possible anisotropy of the variational parameters $R_{\imath}$ and $K_{\imath}$. With this ansatz, the normalization of (\ref{ansatzwigner}) to $N$ fermions leads to
\begin{equation}
48N=\overline{R}^3\overline{K}^3\,, 
\label{particleconservation}
\end{equation}
where the bar denotes geometric averaging, i.e.~$\overline{R}=\left(R_xR_yR_z\right)^{\frac{1}{3}}$. In order to be physically 
self-consistent the ansatz Eq.~(\ref{ansatzwigner}) has 
to minimize the total Hartree-Fock energy of the system as the collision integral vanishes in equilibrium. Hence the total energy 
$E_t$ of the system consists of the kinetic term 
\begin{equation}
 E_{\rm k}=\int d^3r\int \frac{d^3k}{(2\pi)^3}\frac{\hbar^2 {\bf k}^2}{2M}\nu^0({\bf r},{\bf k})\,,  
\label{Ekin} 
\end{equation}
the trapping term 
\begin{equation}
E_{\rm tr}=\int d^3r \int \frac{d^3k}{(2\pi)^3} \frac{M}{2}\left(\sum_{\jmath}\omega_j^2r_j^2\right)\nu^0({\bf r},{\bf k})\,,
\label{Etrapf0}
\end{equation}
the direct Hartree term 
\begin{equation}
E_{\rm d}=\frac{1}{2}\int d^3r\int d^3r' \int \frac{d^3k}{(2\pi)^3}\int \frac{d^3 k'}{(2\pi)^3}V_{\rm d}({\bf r}-{\bf r'})  \nu^0({\bf r},{\bf k})  
\nu^0({\bf r'},{\bf k'})\,, 
\label{Edf0} 
\end{equation}
and the Fock exchange term 
\begin{equation}
E_{\rm ex} =-\frac{1}{2}\int d^3r \int d^3r' \int \frac{d^3k}{(2\pi)^3}\int \frac{d^3 k'}{(2\pi)^3}   V_{\rm d}({\bf r'}) {\rm e}^{i({\bf k}-{\bf k'})
\cdot {\bf r'}} \nu^0({\bf r},{\bf k})  \nu^0({\bf r},{\bf k'})\,. 
\label{Eex}
\end{equation}
Inserting the variational ansatz Eq.~(\ref{ansatzwigner}) for the equilibrium distribution
into the respective energy contributions (\ref{Ekin})--(\ref{Eex}) leads to various phase space integrals.
Whereas both kinetic energy (\ref{Ekin}) and trapping energy (\ref{Etrapf0}) yield elementary solvable integrals,
the computation of the Hartree and Fock integrals (\ref{Edf0}) and (\ref{Eex}) turns out to be more elaborate and is, therefore,
relegated to Appendix A, see also Ref.~\cite{miyakawa_t_2008,Aristeu}. The resulting total energy reads 
\begin{equation}
 E_{\rm t}=\frac{N}{8}\sum_{\jmath}\frac{\hbar^2K_j^2}{2M}+\frac{N}{8}\frac{M}{2}\sum_{\jmath}\omega_j^2R_{\jmath}^2-\frac{48N^2c_0}{8\overline{R}^3}f\left( 
\frac{R_x}{R_z},\frac{R_y}{R_z} 
\right)+\frac{48N^2c_0}{8\overline{R}^3}f\left( \frac{K_z}{K_x},\frac{K_z}{K_y} \right) 
\label{Etotequilib}
\end{equation}
with the constant 
\begin{equation}
c_0=\frac{2^{10}C_{\rm dd}}{3^4\cdot 5\cdot 7\cdot \pi^3}\, . 
\label{C0}
\end{equation}
The Hartree and Fock terms in (\ref{Etotequilib}) depend on the aspect ratio
of the Thomas-Fermi radii and momenta via
the anisotropy function $f(x,y)$, which is defined through the integral
\begin{equation}
f(x,y)=-\frac{1}{4\pi}\int_0^{2\pi}d\phi \int_0^{\pi}d\vartheta \, {\rm sin}\vartheta \left[ \frac{3x^2y^2{\rm cos}^2\vartheta}{\left(y^2{\rm cos}^2 
\phi+x^2{\rm sin}^2\phi\right) {\rm sin}^2\vartheta+x^2y^2{\rm cos}^2\vartheta}-1\right] 
\label{definitionanistropyfunction}
\end{equation}
and which can also be represented as follows \cite{Aristeu,Aristeudoktor}
\begin{equation}
f(x,y)=1 + 3 x y \, \frac{E(\varphi,q) - F(\varphi,q)}{(1-y^2)\sqrt{1-x^2}} \, ,
\end{equation}
where $F(\varphi,q)$ and $E(\varphi,q)$ are the elliptic integrals of the first and second kind, respectively, with $\varphi= \arcsin \sqrt{1-x^2}$
and $q^2=(1-y^2)/(1-x^2)$.
The variational parameters $R_{\imath}$, $K_{\imath}$ are now 
determined by minimizing Eq.~(\ref{Etotequilib}) under the constraint of the particle conservation (\ref{particleconservation}). This leads to the following equations for the momentum parameters $K_{\imath}$
\begin{align}
 {K_x}&={K_y},\label{mom_cyl}\\
  {\frac{\hbar^2 K_z^2}{2M}-\frac{\hbar^2  K_x^2}{2M}}&={\frac{72N c_0}{\overline{R}^3}\left[ 1+\frac{ \left( 2 K_x^2+ K_z^2 \right) f_{\rm s}\left( \frac{K_z}{K_x} \right)}{2\left(  K_z^2-K_x^2  \right)} \right],}
 \label{VSTKz}
\end{align}
with the symmetric anisotropy function $f_{\rm s}(x)=f(x,x)$ \cite{odell,AristeuRapCom,glaumpfau,bosetempdip}, and the spatial parameters $R_{\imath}$

\begin{align}
 &{\omega_x^2R_x^2-\frac{1}{3}\sum_{\jmath}\frac{\hbar^2K_j^2}{M^2}+\frac{48Nc_0}{M\overline{R}^3}\left[ f\left( \frac{R_x}{R_z},\frac{R_y}{R_z}\right)-f_{\rm s}\left(\frac{K_z}{K_x}\right)-\frac{R_x}{R_z} f_1\left(\frac{R_x}{R_z},\frac{R_y}{R_z}\right)  \right]=0,} \label{VSTRx}\\
 &{\omega_y^2R_y^2-\frac{1}{3}\sum_{\jmath}\frac{\hbar^2K_j^2}{M^2}+\frac{48Nc_0}{M\overline{R}^3}\left[ f\left( \frac{R_x}{R_z},\frac{R_y}{R_z}\right)-f_{\rm s}\left(\frac{K_z}{K_x}\right)-\frac{R_y}{R_z} f_2\left(\frac{R_x}{R_z},\frac{R_y}{R_z}\right)  \right]=0, } \label{VSTRy}\\
 &{\omega_z^2R_z^2-\frac{1}{3}\sum_{\jmath}\frac{\hbar^2K_j^2}{M^2}+\frac{48Nc_0}{M\overline{R}^3}\left[ f\left( \frac{R_x}{R_z},\frac{R_y}{R_z}\right)-f_{\rm s}\left(\frac{K_z}{K_x}\right)+\frac{R_x}{R_z} f_1\left(\frac{R_x}{R_z},\frac{R_y}{R_z}\right)+\frac{R_y}{R_z} f_2\left(\frac{R_x}{R_z},\frac{R_y}{R_z}\right)  \right]=0,} \label{VSTRz}
\end{align}
where ${f_{1/2}}$ denotes the derivative of the anisotropy function with respect to first/second argument.

Note that from this result it turns out that the distribution function $\nu^0$ is deformed from a sphere to an ellipsoid
in momentum space due to the Fock term of the dipole-dipole interaction as was first clarified by 
Ref.~\cite{miyakawa_t_2008}. One can see from Eq.~(\ref{mom_cyl}) that the momentum distribution remains cylinder-symmetric even in the case of an anisotropic harmonic trap. Furthermore, a more detailed analysis shows that it always resembles a cigar-shaped form satisfying the inequality ${{K_z>K_x}}$. This follows from the fact that the dipole-dipole interaction (\ref{dipolepotential}) is attractive for head-to-tail orientations, thus stretching the system along the $z$-direction in a prolate cloud shape. The corresponding effect in the particle density had already been obtained before, by means of a Gaussian ansatz in real space 
\cite{rzazewski}. A detailed discussion how the Thomas-Fermi radii and momenta depend on the trap frequencies for both cylinder-symmetric and tri-axial
traps as well as on the dipole-dipole interaction strength $C_{\rm dd}$ can be found, for instance, in 
Refs.~\cite{miyakawa_t_2008,1367-2630-11-5-055017,AristeuRapCom,Aristeu,vladimir}.

On the basis of the variational ansatz Eq.~(\ref{ansatzwigner}) for the equilibrium distribution 
the integrals in Eqs.~(\ref{diffglb}) for the scaling parameters $b_{\imath}$ can now be calculated analytically, yielding
\begin{align}
 \ddot{b}_{\imath}+\omega_i^2b_{\imath}-\frac{\hbar^2 K_{\imath}^2  \Theta_{\imath}}{M^2b_{\imath} R_{\imath}^2 }+ \frac{48N c_0}{ Mb_{\imath} R_{\imath}^2 \prod_j b_j R_{\jmath}}\left[ f\left( \frac{b_xR_x}{b_zR_z},
\frac{b_yR_y}{b_zR_z}\right)-b_{\imath}R_{\imath}\frac{\partial}{\partial b_{\imath} R_{\imath}}f\left( \frac{b_xR_x}{b_zR_z},\frac{b_yR_y}{b_zR_z}\right)\right] \nonumber \\
-\frac{48N c_0}{ Mb_{\imath} R_{\imath}^2 \prod_j b_j R_{\jmath}}\left[ f\left(\frac{\Theta_z^{\frac{1}{2}}K_z}{\Theta_x^{\frac{1}{2}}K_x}, 
\frac{\Theta_z^{\frac{1}{2}}K_z}{\Theta_y^{\frac{1}{2}}K_y}\right)+\Theta_{\imath}^{\frac{1}{2}}K_{\imath} \frac{\partial}{\partial \Theta_{\imath}^{\frac{1}{2}}K_{\imath}} 
f\left(\frac{\Theta_z^{\frac{1}{2}}K_z}{\Theta_x^{\frac{1}{2}}K_x}, \frac{\Theta_z^{\frac{1}{2}}K_z}{\Theta_y^{\frac{1}{2}}K_y}\right)\right]=0\,. 
\label{bequationcomplete}
\end{align}
We remark that our result (\ref{bequationcomplete}) reduces to the corresponding ones obtained in Ref.~\cite{1367-2630-11-5-055017} in the collisionless regime for a cylinder symmetric trapping potential.

Let us now turn to the differential equations for the momentum scaling parameters $\Theta_{\imath}$ in Eq.~(\ref{diffglT}), which still contain the collision integral.
In order to simplify the calculation, we model this collision integral within the widely used relaxation-time approximation
\cite{String,rt1,Dusling,zaremba_book}
\begin{equation}
\label{rta}
I_{\rm coll} [\nu ]\approx -\frac{\nu-\nu^{\rm le}}{\tau}\,.
\end{equation}
Here the phenomenological parameter $\tau$ denotes the relaxation time, which corresponds to the average time between two collisions. Furthermore, we have introduced the local equilibrium 
distribution function 
$\nu^{\rm le}$, which is defined by the condition $I_{\rm coll}[\nu^{\rm le}]=0$ and represents the limiting function of the relaxation process for infinitely large times. The relaxation-time approximation (\ref{rta}) reflects the fact that dissipation should be absent in both HD and CL regimes, where, thus, one expects the collision integral to vanish \cite{String}.

We assume that the collisions only change the 
momentum distribution of $\nu^{\rm le}$ \cite{String}. This is justified by deriving the collision integral within a gradient expansion of the 
distribution function and by considering only the first term  \cite{KadanoffBaym}. Similar arguments have been used before 
in the context of the local density approximation for bosonic dipolar gases \cite{quantumfluctuations,beyondmeanfield}. Therefore, 
$\nu^{\rm le}$ is determined from rescaling the equilibrium distribution  $\nu^0$ 
via an ansatz similar to Eq.~(\ref{scalingWigner}), i.e.
\begin{equation}
\nu^{\rm le}({\bf x},{\bf q},t)=\Gamma^{\rm le} \nu^0({\bf r}(t),{\bf k}^{\rm le} (t)) \, , 
\label{localscaling}
\end{equation}
with the old scaling parameters $b_{\imath}$ in real space according to (\ref{b}), but new scaling parameters $\Theta_{\imath}^{\rm le}$ in momentum space
\begin{equation}
k_i^{\rm le} =\frac{1}{\sqrt{\Theta_{\imath}^{\rm le} (t)}}\left[ q_i-\frac{M\dot{b}_{\imath}(t) x_i}{\hbar b_{\imath}(t)}\right]\,,
\label{scalingwignerkle}
\end{equation}
thus yielding the corresponding normalization 
\begin{equation}
\label{con}
\Gamma^{\rm le}=\frac{1}{\prod_j b_j \sqrt{\Theta_{\jmath}^{\rm le}} }  \,.
\end{equation}
Inserting the ansatz (\ref{rta}) into Eq.~(\ref{diffglT}) finally leads to
\begin{equation}
\dot{\Theta}_i+2\frac{\dot{b}_{\imath}}{b_{\imath}}\Theta_{\imath}=-\frac{1}{\tau}\left( \Theta_{\imath}-\Theta_{\imath}^{\rm le}\right) \, .
\label{thetaequationcomplete}
\end{equation}
The physical meaning of this equation is that dissipation occurs in the system outside of local equilibrium in each direction separately as long as there are collisions, i.e., as long as the relaxation time $\tau$ remains finite. As a matter of fact, in order to obtain a closed set of equations we have yet to find additional equations which determine the momentum scaling parameters $\Theta_{\imath}^{\rm le}$. In the case of a Fermi gas with contact interaction only, a relation between the scaling in local equilibrium $\Theta_{\imath}^{\rm le}$ for different directions can be obtained as in Ref.~\cite{String} and they turn out to be all equal, due to the isotropy of the system. For a dipolar gas, however, this is not valid anymore and a different approach must be followed. Indeed, due to the presence of the DDI, the Fermi sphere is deformed, thus reducing the symmetry in the momentum scaling from spherical to ellipsoidal. To overcome this issue, we build on the fact that a hydrodynamic theory was already worked out in Refs.~\cite{AristeuRapCom,Aristeu} in which local equilibrium is always ensured \cite{String}. To this end, we study the relation between the momentum scaling parameters in the framework of the hydrodynamic theory and evaluate the respective energy contributions Eqs.~(\ref{Ekin})--(\ref{Eex}) in local equilibrium in (\ref{localscaling}), where again the collision integral vanishes by definition. With this we obtain 
\begin{align}
E_{\rm t}^{\rm le}=&-\frac{N}{8}\frac{M}{2}\sum_i \dot{b}_{\imath}^2 R_{\imath}^2+\frac{N}{8}\sum_i \frac{\hbar^2 K_{\imath}^2 \Theta_{\imath}^{\rm le}}{2M}+\frac{N}{8} \frac{M}{2} \sum_i 
\omega_i^2 b_{\imath}^2 R_{\imath}^2-\frac{48N^2c_0}{8\prod_j b_j R_{\jmath}}f\left( \frac{b_x R_x}{b_z R_z},\frac{b_y R_y}{b_z R_z} \right)\nonumber \\
&+\frac{48N^2c_0}{8\prod_j b_j R_{\jmath}}f\left( \frac{(\Theta_z^{\rm le})^{\frac{1}{2}} K_z}{(\Theta_x^{\rm le})^{\frac{1}{2}} K_x},
\frac{(\Theta_z^{\rm le})^{\frac{1}{2}} K_z}{(\Theta_y^{\rm le})^{\frac{1}{2}} K_y} \right)\,. 
\label{Etotloc}
\end{align}
However, when determining the momentum scaling parameters $\Theta_{\imath}^{\rm le}$ by minimizing Eq.~(\ref{Etotloc}), we have to consider that they 
turn out to be not independent of one another. Summing all three Eqs.~(\ref{thetaequationcomplete}) yields in local equilibrium the constraint
\begin{equation}
\label{conn}
\prod_j b_j \sqrt{\Theta_{\jmath}^{\rm le}} =1  \,.
\end{equation}
With this the minimization of the energy (\ref{Etotloc}) leads to
\begin{align}
\Theta_x^{\rm le}&=\Theta_y^{\rm le}\, ,
\label{thetaxthetay} \\
\frac{\hbar^2 \Theta_z^{\rm le} K_z^2}{2M}-\frac{\hbar^2 \Theta_x^{\rm le} K_x^2}{2M}&=\frac{3}{2}\frac{48N c_0}{\prod_j b_j R_{\jmath}}\left\{ 1+\frac{ 
\left( 2\Theta_x^{\rm le} K_x^2+\Theta_z^{\rm le} K_z^2 \right) f_{\rm s}\left[ \frac{\left(\Theta_z^{\rm le}\right)^{\frac{1}{2}} 
K_z}{\left(\Theta_x^{\rm le}\right)^{\frac{1}{2}} K_x} \right]}{2\left( \Theta_z^{\rm le} K_z^2-\Theta_x^{\rm le} K_x^2\right)} \right\}.
\label{couplinglocalequ}
\end{align}
These two equations show the deformation of the local equilibrium distribution function in momentum space, which disappears if we set the dipole-dipole interaction to zero. Again we obtain a cylinder-symmetric configuration in momentum space satisfying {${\sqrt{\Theta^{\rm le}_z}K_z>\sqrt{\Theta^{\rm le}_x}K_x}$}. Furthermore, we note that 
the right-hand side of Eq.~(\ref{couplinglocalequ}) originates from the Fock term, hence the momentum distribution remains 
cylinder-symmetric even in 
the case of an anisotropic harmonic trap. In the absence of the DDI, i.e. more precisely in the absence of the Fock exchange term of the DDI, 
the momentum scaling parameters in local equilibrium assume the same values 
in all three directions. This resembles the case of a two-component Fermi gas featuring contact interaction only \cite{String}.

Finally, we remark that a solution in the hydrodynamic regime, where the relaxation time goes to zero, i.e.~$\tau \rightarrow 0$, must have the same momentum symmetry as in the case of local equilibrium. This is a common feature in both cases of contact and dipolar interactions. Thus, the hydrodynamic theory in Refs.~\cite{AristeuRapCom,Aristeu} turns out to be crucial for the determination of the momentum scaling parameters in this regime, as given by Eqs.~(\ref{conn})--(\ref{couplinglocalequ}).

\section{Low-Lying Excitation Modes}\label{linear}

Having solved the Hartree-Fock Boltzmann-Vlasov theory with a scaling ansatz involving the local equilibrium solution, which is properly linked to the hydrodynamic regime, we can address various dynamical properties of interest. Whereas the time-of-flight dynamics of a dipolar Fermi gas has recently been studied along the same lines developed in the present article \cite{vladimir}, here we analyze the low-lying excitations of the system. To this end, we use the equations of motion (\ref{bequationcomplete}), (\ref{thetaequationcomplete}), and (\ref{conn})--(\ref{couplinglocalequ}) for the scaling parameters  $b_{\imath}$, $\Theta_{\imath}$, and $\Theta_{\imath}^{\rm le}$ and calculate the various properties of the low-lying excitation modes via a linearization around the respective equilibrium values.
\subsection{Linearization}

In order to obtain the low-lying excitation frequencies, one has to linearize the equations of motion for the scaling parameters. To this end, we decompose all spatial and momentum scaling parameters according to 
\begin{equation}
b_{\imath}=b_{\imath}^0+\delta b_{\imath},\quad\Theta_{\imath}=\Theta_{\imath}^0+\delta \Theta_{\imath}, \quad\Theta_{\imath}^{\rm le}=\Theta_{\imath}^{{\rm le},0}+\delta \Theta_{\imath}^{\rm le},
\label{linear1}
\end{equation}
with the equilibrium values $b_{\imath}^0=\Theta_{\imath}^0=\Theta_{\imath}^{{\rm le},0}=1$ for all $\imath$. This decomposition leads to cumbersome equations for the respective elongations $\delta b_{\imath}$, $\delta \Theta_{\imath}$, and $\delta \Theta_{\imath}^{\rm le}$ out of equilibrium, which are relegated to the Appendix \ref{linearization}. For now, it suffices to consider the following physical aspects.

The first aspect we would like to point out is the cylinder symmetry of the momentum distribution which implies for local equilibrium $\delta \Theta_x^{\rm le}- \delta \Theta_y^{\rm le}=0$, as follows directly from the linearization process in (\ref{linequ1}). This can be traced back to the anisotropy of the DDI. Indeed, the momentum distribution of a Fermi gas at low temperatures is dominated by the Fermi pressure, which is isotropic. In turn, the DDI singles out the $z$-direction, as the dipoles are aligned along that direction. This is in contrast to the case of Bose gases at zero temperature, since the occupation of a single one-particle mode suppresses the Fock term \cite{Baillie2} and the momentum distribution is dominated by the anisotropy in real space. For finite temperature Bose gases, on the other hand, the momentum distribution is correspondingly deformed \cite{Baillie2}.

Proceeding with the linearization, one obtains with (\ref{thetaxle}) and (\ref{thetayle}) a set of two equations relating the elongations in momentum space in local equilibrium $\delta \Theta_x^{\rm le}, \delta \Theta_z^{\rm le}$ to the elongations in real space $\delta b_{\imath}$ in all three directions $j=x,y,z$. The remaining six parameters $\delta b_{\imath}$ and $\delta \Theta_{\imath}$ are determined by (\ref{iso_mom_delta})--(\ref{deltadotdotb}).

In order to apply the theory developed here, let us restrict our calculations to the cylinder-symmetric case, which is capable of displaying the physical properties accessible in actual experiments. Thus, from now on, we adopt a cylinder-symmetric trapping potential with $\omega_x=\omega_y=\omega_{\rho}$ and $\omega_z=\lambda \omega_{\rho}$, where $\lambda$ denotes the trap aspect ratio. To this end, we use
\begin{equation}
\lim_{y \rightarrow x}{xf_1(x,y)}=\lim_{y \rightarrow x}{yf_2(x,y)}=-1+\frac{(2+x^2)f_s(x)}{2(1-x^2)},
\end{equation}
so that the derivatives of the anisotropy function can be re-expressed as algebraic functions containing $f_s$. This function first appeared in the study of dipolar BECs with a Gaussian density distribution \cite{liyou} and has been found to be characteristic of the DDI for dipolar quantum gases. For example, it occurs in the case of an exact Thomas-Fermi solution of the Gross-Pitaevskii equation \cite{eberlein} and in the context of determining the DDI shift of the BEC critical temperature \cite{glaumpfau,bosetempdip}.

The differential equations relating the momentum and real space elongations according to (\ref{iso_mom_delta})--(\ref{deltadotdotb}) can be solved by assuming that all deviations from equilibrium oscillate with one and the same frequency $\Omega$:
\begin{equation}
\delta b_{\imath}=\xi_{\imath} {\rm e}^{i \Omega t}\,,\hspace{1cm}
\delta \Theta_{\imath}=\chi_{\imath} {\rm e}^{i \Omega t}\,.
\label{PW}
\end{equation}

Due to the inclusion of the collisional term within the relaxation-time approximation, we obtain complex frequencies $\Omega$, whose real parts represent the eigenfrequencies of the low-lying modes of the system and whose imaginary parts describe the corresponding damping rates. Simple algebraic manipulations allow due to (\ref{iso_mom_delta})--(\ref{deltadotthetay}) at first to re-express the amplitudes in momentum space ${\chi_{\imath}}$ as functions of the ones in real space via

\begin{equation}
\chi_{\imath}=\frac{\tau}{1+i \Omega \tau}\left( -2i \Omega \xi_{\imath}+\frac{g_{\imath}}{\tau}\sum_{\jmath} \xi_{\jmath} \right), \label{amplitudesrealmomentum}
\end{equation}
where the coefficients $g_{\imath}$ are given by
\begin{equation}
g_{x} = \frac{A-2C}{B+2C},\quad g_{z} = -2 \frac{A + B}{B+2C}.
\end{equation}
This equation (\ref{amplitudesrealmomentum}) allows to deduce directly from the amplitudes of the oscillations in real space the corresponding amplitudes in momentum space. Inserting Eq.~(\ref{amplitudesrealmomentum}) into (\ref{deltadotdotb}) we then obtain three coupled equations for the amplitudes $\xi_{\imath}$ in real space
\begin{equation}
-\Omega^2 \xi_{\imath}+\sum_{\jmath}\left( O_{\imath \jmath}-\frac{2i \Omega \tau}{1+i \Omega \tau}D_{\imath \jmath}+\frac{\alpha_i}{1+i \Omega \tau}\right) \xi_{\jmath}=0\,,
\label{LA}
\end{equation}
where we have introduced the abbreviations
\begin{eqnarray}
\alpha_x&=& \frac{A-2C}{B+2C} (D_{xx}+D_{xy})-2\frac{A+B}{B+2C}D_{xz}\,,\\ 
\alpha_z&=& 2\frac{A-2C}{B+2C} D_{zx}-2\frac{A+B}{B+2C}D_{zz}\,,
\end{eqnarray}
while the abbreviations $A, B, C, O_{\imath \jmath}\textrm{, and } D_{\imath \jmath}$ are defined in the Appendix \ref{linearization} as functions of the Thomas-Fermi radii and momenta.

The key feature here is to notice that Eq. (\ref{LA}) has nontrivial solutions, provided that the corresponding determinant vanishes:
\begin{eqnarray}
0&=&\left[ O_{xx}-O_{xy}-\frac{2i \Omega \tau}{1+i \Omega \tau}\left(D_{xx}-D_{xy} \right)-\Omega^2 \right] \left\{ -2\left( O_{xx}
-\frac{2i \Omega \tau D_{xz}-\alpha_x}{1+i \Omega \tau} \right) \left( O_{zz}-\frac{2i \Omega \tau D_{zx}-\alpha_z}{1+i \Omega \tau} \right) \right. \nonumber \\
&&\left.+\left[ O_{xx}+O_{xy}-\frac{2i \Omega \tau \left( D_{xx}+D_{xy}\right)-2\alpha_x}{1+i \Omega \tau}-\Omega^2 \right] 
\left( O_{zz}-\frac{2i \Omega \tau D_{zz}-\alpha_z}{1+i \Omega \tau} -\Omega^2 \right)\right\}\,. \label{determinantequation}
\end{eqnarray}
This can be put in a compact form which is similar to one for a Bose gas with contact interaction \cite{String}
\begin{equation}
\left(P[\Omega]+\frac{1}{i \Omega \tau}Q[\Omega]\right) \left(S[\Omega]+\frac{1}{i \Omega \tau}T[\Omega]\right)=0\,, 
\label{determinant1}
\end{equation}
where the respective polynomials are given by 
\begin{eqnarray}
P[\Omega]&=&(\Omega^2-\Omega_{+;{\rm CL}}^2)(\Omega^2-\Omega_{-;{\rm CL}}^2)\,, \hspace{0.5cm}
Q[\Omega]=(\Omega^2-\Omega_{+;{\rm HD}}^2)(\Omega^2-\Omega_{-;{\rm HD}}^2)\,,\\ 
S[\Omega]&=&\Omega^2-\Omega_{\rm rq;CL}^2,  \hspace{0.5cm} T[\Omega]=\Omega^2-\Omega_{\rm rq;HD}^2\, .
\end{eqnarray}
The explicit expressions for the low-lying oscillation frequencies can be obtained for the limiting cases in which the relaxation time either diverges, i.e. in the collisionless regime 
\cite{miyakawa_t_2008,1367-2630-11-5-055017,PhysRevA.83.053628} 
\begin{eqnarray}
\Omega_{\rm rq;CL}^2&=&O_{xx}-O_{xy}+2\left( D_{xy}-D_{xx} \right)\,,\label{OmegarqCL}\\
\Omega_{+;{\rm CL}}^2&=&\frac{1}{2}\left( -2D_{xx}-2D_{xy}-2D_{zz}+O_{xx}+O_{xy}+O_{zz}+ \sqrt{R_1} \,\right) \,,\label{OmegamoCL}\\
\Omega_{-;{\rm CL}}^2&=&\frac{1}{2}\left( -2D_{xx}-2D_{xy}-2D_{zz}+O_{xx}+O_{xy}+O_{zz}- \sqrt{R_1} \,\right) \,,\label{OmegatqCL}
\end{eqnarray}
or vanishes, i.e. in the hydrodynamic regime \cite{AristeuRapCom,Aristeu}
\begin{eqnarray}
\Omega_{\rm rq;HD}^2&=&O_{xx}-O_{xy}\,,\label{Omegarqhy}\\
\Omega_{+;{\rm HD}}^2&=&\frac{1}{2}\left(2\alpha_x+\alpha_z+O_{xx}+O_{xy}+O_{zz}+\sqrt{R_2}\, \right)\,, \label{Omegamohy}\\
\Omega_{-;{\rm HD}}^2&=&\frac{1}{2}\left(2\alpha_x+\alpha_z+O_{xx}+O_{xy}+O_{zz}-\sqrt{R_2}\, \right)\,. \label{Omegatqhy}
\end{eqnarray}
Here the subscribes (rq), $(+)$, $(-)$ denote the radial quadrupole mode, the monopole mode, and the three-dimensional quadrupole 
mode, respectively, and we have introduced the additional abbreviations
\begin{eqnarray}
R_1&=&\left( 2D_{xx}+2D_{xy}+2D_{zz}-O_{xx}-O_{xy}-O_{zz} \right)^2-4\left( -8D_{xz}D_{zx}+4D_{xx}D_{zz}+4D_{xy}D_{zz}-2D_{zz}O_{xx}\right. \nonumber \\
&&\left.-2D_{zz}O_{xy}+4D_{zx}O_{xz}+4D_{xz}O_{zx}-2O_{xz}O_{zx}-2D_{xx}O_{zz}-2D_{xy}O_{zz}+O_{xx}O_{zz}+O_{xy}O_{zz} \right), \\
R_2&=&\left( 2\alpha_x+\alpha_z+O_{xx}+O_{xy}+O_{zz} \right)^2-4\left( \alpha_zO_{xx}+\alpha_zO_{xy}-2\alpha_zO_{xz}\right.\nonumber \\
&&\left.-2\alpha_xO_{zx}-2O_{xz}O_{zx}+2\alpha_xO_{zz}+O_{xx}O_{zz}+O_{xy}O_{zz} \right)\,.
\end{eqnarray}

We conclude that Eq.~(\ref{determinant1}) represents the main result of the present article: it allows to obtain both the excitation frequencies and the corresponding damping rates of the low-lying excitation of all three low-lying excitation modes of a dipolar Fermi gas in a triaxial harmonic trap all the way from the collisionless to the hydrodynamic regime. Thereby, the special limits in the HD and CL regimes are made explicit. Our main result (\ref{determinant1}) opens the road to study low-lying collective frequencies in the different collisional regimes, which should be measurable in the near future.
\subsection{Collisionless and Hydrodynamic Regime}
Although it is not possible to determine analytically from (\ref{determinant1}) the complex frequencies $\Omega$ of the low-lying excitation modes for an arbitrary relaxation time $\tau$, explicit results have been obtained for the limiting cases for the monopole $\Omega_{+}$, the three-dimensional quadrupole $\Omega_{-}$, and the radial quadrupole $\Omega_{\rm rq}$ frequencies, according to Eqs.~(\ref{OmegarqCL})--(\ref{Omegatqhy}). In the present section we gain some physical insight in light of these analytical expressions and the corresponding graphs featuring the dependence of the collective excitation frequencies on the DDI for different types of trapping potentials.

In order to set the stage for the discussion of the physical properties of the system, let us introduce dimensionless variables by referring to the noninteracting case, which provides adequate units for all quantities of interest such as the Thomas-Fermi radius $R_{\imath}^0=\sqrt{\displaystyle {2E_F}/{M\omega_i^2}}$ and the corresponding Fermi momentum $K_F=\sqrt{\displaystyle {2ME_F}/{\hbar^2}}\,$. The latter, in turn, depends on the Fermi energy $E_F=\hbar \overline{\omega}(6N)^{\frac{1}{3}}$. Here, the geometric mean of the trap frequencies reads $\overline{\omega}=\omega_\rho \lambda^{1/3}$, where $\lambda= \omega_{z}/\omega_{\rho}$ denotes the trap aspect ratio. Using these physical dimensions leads to the dimensionless 
dipole-dipole interaction strength 
\begin{equation}
\epsilon_{\rm dd} =\frac{C_{\rm dd}}{4\pi}\left( \frac{M^3 \overline{\omega}}{\hbar^5} \right)^{\frac{1}{2}}N^{\frac{1}{6}} \, .
\end{equation}
\setlength{\unitlength}{1cm}
\begin{center}
\begin{figure}[t]
\includegraphics[scale=0.98]{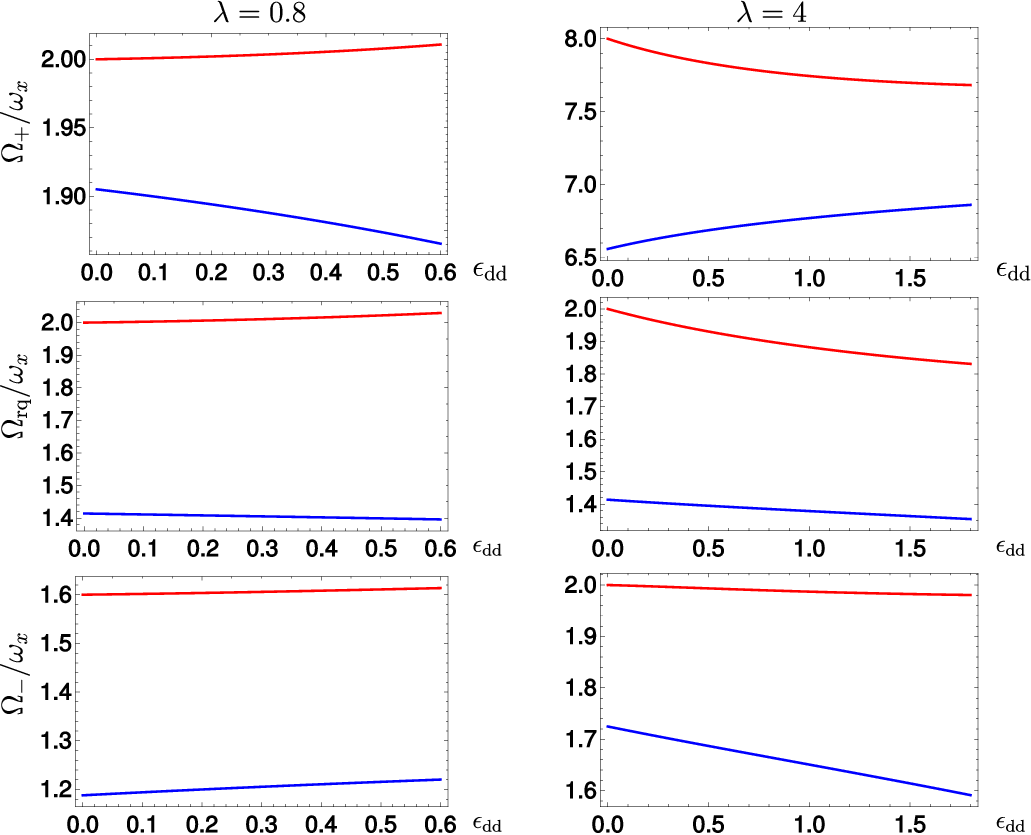}
\caption{(Color on-line) Low-lying collective excitation frequencies within a cylinder-symmetric trap for the monopole ($\Omega_{+}$), the three-dimensional quadrupole ($\Omega_{-}$), and the radial quadrupole modes ($\Omega_{\rm rq}$). They are presented in both the collisionless regime
(upper red curves) and the hydrodynamic regime
(lower blue curves) as a function
of the dimensionless dipolar interaction strength $\epsilon_{\rm dd}$ for cigar-shaped ($\lambda < 1$) and pancake-shaped ($\lambda > 1$) traps.}
\label{fig1}
\end{figure}
\end{center}

At first, we discuss the frequencies of all three low-lying modes in the limiting cases of the collisionless regime ($\tau \rightarrow \infty$) and the hydrodynamic regime ($\tau \rightarrow 0$), which are depicted in Fig.~\ref{fig1} as a function of $\epsilon_{\rm dd}$ for the trap aspect ratios $\lambda=0.8$ and $4$. We observe at first that the collisionless frequencies turn out to be always larger than the corresponding hydrodynamic frequencies. This can be most illustratively explained in the case of the radial quadrupole mode. The eigenvector in real space $(\xi_1,\xi_2,\xi_3)$ takes the general form $(1,-1,0)$ so the mode oscillates in $x$- and $y$-direction out of phase and no oscillation occurs in $z$-direction. The corresponding amplitudes in momentum space $\chi_{\imath}$ turn out to vanish in the hydrodynamic regime. This means that no oscillation in $k$-space occurs in the hydrodynamic limit, which stems from the fact that the high collision rate ensures local equilibrium. In contrast, in the collisionless regime the amplitudes in momentum space $\chi_{\imath}$ turn out to have values with opposite sign to the amplitudes in real space $\xi_{\imath}$. Thus, intuitively, since the collisionless oscillation involves both real and momentum amplitudes, one expects it to have a higher energy than in the hydrodynamic regime, which involves only real amplitudes. Indeed, this can be further analyzed with the frequencies $\Omega_{\rm rq;HD}$ and $\Omega_{\rm rq;CL}$ from (\ref{OmegarqCL}) and (\ref{Omegarqhy}), which reduce to the concise expressions
\begin{align}
\frac{\Omega_{\rm rq;HD}^2}{\omega_x^2}=&2+\frac{3 \lambda^2 \epsilon_{\rm dd} c_{\rm d}}{4 \prod_j \tilde{R}_j}\frac{2\left( \tilde{R}_z^2-\lambda^2 
\tilde{R}_x^2\right)-\left(4\tilde{R}_z^2+\lambda^2\tilde{R}_x^2\right) 
f_s\left(\frac{\lambda\tilde{R}_x}{\tilde{R}_z}\right)}{\left(\tilde{R}_z^2-\lambda^2\tilde{R}_x^2\right)^2} \, ,\label{RQHY}\\
\frac{\Omega_{\rm rq;cl}^2}{\omega_x^2}=&\frac{\Omega_{\rm rq;HD}^2}{\omega_x^2}
 +2\frac{\tilde{K}_x^2}{\tilde{R}_x^2}+\frac{\tilde{K}_z^2}{\tilde{R}_x^2} \frac{\left(4+\frac{\tilde{K}_z^2}{\tilde{K}_x^2}\right)\left(2+\frac{\tilde{K}_z^2}{\tilde{K}_x^2}\right)f_s\left(\frac{\tilde{K}_z}{\tilde{K}_x}\right)+4\left(1-\frac{\tilde{K}_z^2}{\tilde{K}_x^2}\right)}{2\left(2+\frac{\tilde{K}_z^2}{\tilde{K}_x^2}\right)\left[\left(2+\frac{\tilde{K}_z^2}{\tilde{K}_x^2}\right)f_s\left(\frac{\tilde{K}_z}{\tilde{K}_x}\right)-2\left(1-\frac{\tilde{K}_z^2}{\tilde{K}_x^2}\right)\right]} 
\, ,\label{RQCL}
\end{align}
where the dimensionless Thomas-Fermi radii and momenta read $\tilde{R}_i=R_{\imath}/R_{\imath}^{0}$ and $\tilde{K}_i=K_{\imath}/K_F$ and we have introduced the constant $c_d={2^{\frac{38}{3}}}/{3^{\frac{23}{6}}\cdot 5 \cdot 7 \cdot \pi^2}$. Note, however, that our result (\ref{RQHY}), (\ref{RQCL}) differs
significantly from the corresponding one of Ref.~\cite{String}, where the Fock exchange term and, thus, the deformation of the Fermi sphere to an ellipsoid is not
taken into account.
Note that (\ref{RQCL}) reveals explicitly that the radial quadrupole mode frequency in the collisionless regime is larger than in the hydrodynamic regime. The first additional term  corresponds to twice the kinetic energy in momentum space and the second additional term is due to the deformation of the Fermi sphere and turns out to be always positive.

Physically, the frequencies of all the three modes in the hydrodynamic regime are expected to be lower due to the fact that the kinetic energy is negligible in this regime in comparison to the collisionless case. This analysis shows an agreement of the present case of a dipolar Fermi gas with the Bose gas with contact interaction only, as discussed in Ref.~\cite{String}, in the sense that collisionless frequencies are usually higher than the hydrodynamic ones.

We would like to remark that our results for the three-dimensional modes, in both regimes, and for the radial quadrupole mode in the HD regime and with dipolar induced Fermi sphere deformation do agree with the ones previously found in literature \cite{1367-2630-11-5-055017,AristeuRapCom,Aristeu}.

Furthermore, the overall picture that the frequencies in both regimes tend to coincide as the kinetic energy becomes negligible in comparison with the mean-field contribution, found for Bose gases with contact interaction \cite{String}, is also seen here, but only for very oblate traps. Consider, for instance, the case of a trap with $\lambda=4$, as shown in the right column of Fig. \ref{fig1}. In the hydrodynamic regime of a dipolar Fermi gas, the excitation frequency of the monopole mode for a gas in a pancake like trap increases \cite{AristeuRapCom,Aristeu} and could, in principle, reach the higher value assumed in the collisionless regime, as indicated by the upper right plot. These values will not, however, coincide for any value of $\epsilon_{\rm dd}$ due to the fact that the DDI is partially attractive and increasing $\epsilon_{\rm dd}$ will necessarily lead to a collapse of the system. Moreover, the values of the other two frequencies decrease and actually vanish, as the dipolar interaction strength $\epsilon_{\rm dd}$ reaches a given threshold value \cite{AristeuRapCom,Aristeu}. For prolate traps, which are more vulnerable to collapse as they favor attraction between the dipoles, such a convergence between the frequencies in collisionless and hydrodynamic regimes does not appear at all, as one can see in the left column of Fig. \ref{fig1}.

\subsection{From Collisionless to Hydrodynamic Regime}

Let us now turn to the interpolation between the two limiting regimes. This can be studied by numerically solving Eq.~(\ref{determinant1}) for different values of the relaxation time $\tau$, ranging all the way from very low values, corresponding to the HD regime, to very large ones, which correspond to the hydrodynamic regime.

\begin{center}
\begin{figure}[t]
\includegraphics[scale=1.0]{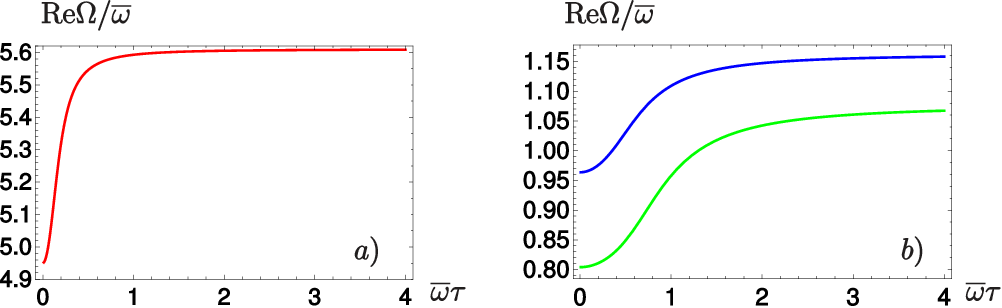}
\caption{(Color on-line) Low-lying excitation mode frequencies within a cylinder-symmetric trap with respect to the relaxation time $\tau$ 
for the trap aspect ratio $\lambda=5$ and the dimensionless interaction strength $\epsilon_{\rm dd}=1.33$;
a) monopole mode (red curve), b) radial quadrupole (lower green curve) and three-dimensional quadrupole mode (upper blue curve).}
\label{fig2}
\end{figure}
\end{center}

As a matter of fact, we have found that the excitation frequencies of the low-lying modes exhibit a characteristic qualitative dependence on the relaxation time $\tau$, even when the trap anisotropy $\lambda$ and the DDI strength $\epsilon_{\rm dd}$ take a wide range of values. Let us, therefore, consider exemplarily the frequencies of the low-lying modes for the trap anisotropy parameter $\lambda= 5$, which corresponds to a  pancake-like cloud, and for the relative dipolar strength $\epsilon_{\rm dd}=1.33$ as shown in Fig.~\ref{fig2}. 
\setlength{\unitlength}{1cm}
According to the previous subsection, we have observed that the frequencies of all eigenmodes have smaller values in the HD than in the CL regime. Moreover, they increase monotonously with the relaxation time and, eventually, reach a plateau for larger values of the relaxation time, in which the system can be well described as completely CL. Despite these qualitative features, which are common to all three modes, it is interesting to remark that the passage from the HD to the CL regime with increasing 
relaxation time occurs differently for the respective modes. Indeed, comparing the graphs in Fig.~\ref{fig2}, we see that the transition from the HD to the CL regime with increasing relaxation time $\tau$ is most abrupt for the monopole mode, while being quite smooth for the radial quadrupole mode. As for the three-dimensional quadrupole, the transition happens in an intermediate way between the other two. 

From an experimental point of view, this behavior has important consequences. In fact, an actual sample with given dipolar strength, particle number, and trap frequencies would correspond a given relaxation time. Take, for example, the value or the relaxation time for which $\overline{\omega}\tau = 0.8$ holds, with the mean harmonic frequency $\overline{\omega}$. For the monopole mode, on the one hand, there would be nearly no distinction between the frequency value and the one in the CL limit ($\overline{\omega}\tau \rightarrow\infty$). For the other two modes, on the other hand, one would obtain frequency values, which correspond to the transition region between the two regimes.
\begin{figure}[t]
\includegraphics[scale=0.9]{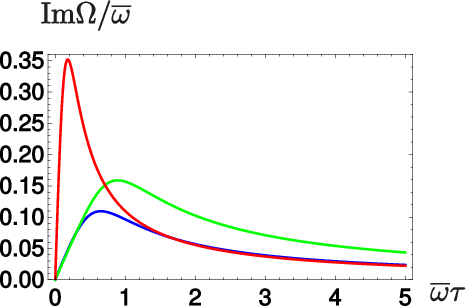}
\caption{(Color on-line) Damping of monopole mode (upper red curve), radial-quadrupole mode (middle green curve), and three-dimensional quadrupole mode (lower blue curve) 
plotted with the same values for the trap aspect ratio and the dimensionless interaction strength as in Fig.~\ref{fig2}.}
\label{fig3}
\end{figure}

This overall picture of how the transition from the collisional to the hydrodynamic regime occurs is confirmed if one also analyzes the imaginary parts of the complex frequencies $\Omega$, which represent the damping rates of the low-lying collective modes. First of all, we note that they vanish in the limiting cases of the hydrodynamic and collisionless regime, as depicted 
in Fig.~\ref{fig3}. This is compatible with the fact that dissipation occurs neither in the HD nor in the CL regime. Furthermore, from comparing the real with the imaginary parts of the frequencies, we read off two important conclusions. 

At first, both the position and the width of the damping peaks 
reveal the respective regions, where the main change of the real part of the frequencies occurs. 
This enables to determine the crossover regions as well as 
the regions in which the system behaves mainly either hydrodynamic or collisionless. Accordingly, one recognizes in Fig.~3
that the transition for the monopole mode is at first most abrupt, i.e.
the oscillation frequency changes by a large amount, while the ones for the other two modes take place for larger values of the 
relaxation times and their frequency change is not so large. We offer an interpretation of this fact based on the observation that the monopole mode is the only one in which the oscillations in the different directions are in phase with one another, i.e., the cloud is either compressed or expanded in all three directions at the same time, therefore this mode is also called the breathing mode. Consider, for example, the period in which the gas is compressed. Since collisions affect the way in which the cloud can be compressed, one should expect them to have a larger impact when the compression occurs in all three directions at the same time, in comparison with cases in which a compression in one direction takes place simultaneously with an expansion in other directions. Thus, according to this reasoning, the breathing mode should, indeed, be more sensible to the collisional regime than the other two. And this explains why the transition from the CL to the HD regime is remarkably different for this mode.

\begin{figure}[t]
\includegraphics[scale=0.9]{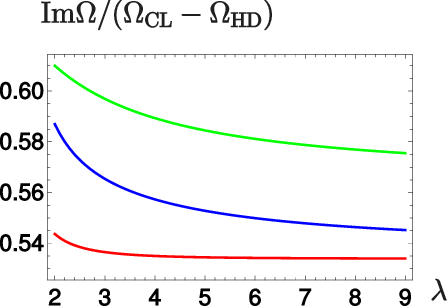}
\caption{(Color on-line) Ratio of dissipative peak height and difference of collisionless and hydrodynamic frequency for the radial quadrupole mode (green upper curve)
according to (\ref{peakhightanalytic}), the three-dimensional quadrupole mode (blue middle curve), and the monopole mode (red lower curve)
as a function of the trap aspect ratio $\lambda$ at the dimensionless interaction strength $\epsilon_{\rm dd}=1.33$.}
\label{fig4}
\end{figure}

The second conclusion is that the damping of the oscillations exhibits a peak for an intermediate relaxation time. 
A quantitative analysis reveals that the height of this peak is in good approximation proportional to the difference between the real parts of the frequencies in the hydrodynamic and the collisionless regime \cite{Stringari}. However, detailed numerical studies show small deviations from this general behavior. Therefore, we analyzed the dependence of the peak height from the limiting frequencies analytically for the radial quadrupole mode, whose frequency follows from (\ref{determinant1}). Splitting the complex frequency $\Omega$ into its respective 
real and imaginary part allows to derive an analytic formula for the peak height, which turns out to depend 
on the limiting frequencies as follows
\begin{equation}
{\rm Im} \Omega(\tau^*)=\frac{1}{4}\left( \frac{\Omega_{\rm rq;CL}}{\Omega_{\rm rq;HD}}+1 \right)
\left( \Omega_{\rm rq;HD}-\Omega_{\rm rq;CL} \right)\,, 
\label{peakhightanalytic}
\end{equation}
where $\tau^*$ denotes the relaxation time at the peak in the imaginary part, which is determined from
$\frac{d {\rm Im} \Omega(\tau)}{d\tau}|_{\tau=\tau^*}=0$. Equation (\ref{peakhightanalytic}) shows that the peak height of the imaginary part of the radial quadrupole mode changes approximately linear with the difference between the limiting frequencies. This represents an important result of our analysis as it allows to infer collisionless and hydrodynamic frequencies of the low-lying modes from measuring the maximal damping. A similar procedure allows for homogeneous Fermi gases to determine the velocities of zero sound \cite{VollhardtWoelfle}. However, the prefactor in Eq.~(\ref{peakhightanalytic}) leads to a small deviation from this linear dependence for the radial quadrupole mode.
In order to reveal this deviation graphically, one has to choose a large value for the relative dipolar interaction strength 
$\epsilon_{\rm dd}$, which, however, excludes a cigar-like cloud due to the instability of the dipolar interaction \cite{AristeuRapCom,Aristeu}.
According to Fig.~\ref{fig4} the ratio of dissipative peak height and difference of collisionless and hydrodynamic frequencies decreases 5.7 \%, once the
trap aspect ratio $\lambda$ increases from 2 to 9 for  $\epsilon_{\rm dd}=1.33$.
Numerically we find that also the other two modes reveal a similar small deviation from the 
linear dependence of the imaginary peak height from the difference of the limiting frequencies, which turns out to be most pronounced for the three-dimensional
quadrupole mode.

\subsection{Experimental Prospects}

Let us now consider the experimental prospects for the observation of the collisional properties of dipolar Fermi gases. We have presented quantitative results for the low-lying excitations of Fermi gases, ranging all the way from the collisionless to the hydrodynamic regimes. All these detailed predictions warrant a quantitative comparison with experimental results. To this end, there are some quite promising candidates. For instance, the fermionic isotope of $^{53}$Cr \cite{Chromium}, which has a magnetic moment of $m=6$ Bohr magnetons, the fermionic isotope of $^{167}$Er \cite{Ferlaino,ferlaino_deform} with $m=7$ Bohr magnetons, or the fermionic isotope of $^{161}$Dy \cite{PhysRevLett.108.215301} which has an even larger magnetic moment of $m=10$ Bohr magnetons. Among these, $^{167}$Er \cite{Ferlaino,ferlaino_deform} seems to be most adequate to experimentally detect the effect of the Fermi sphere deformation in terms of the low-lying excitation modes. Indeed, time-of-flight experiments have been investigated with the help of a theoretical background developed along the lines of the present article \cite{vladimir}. In fact, it has been found by means of a time-of-flight analysis that this deformation in momentum space is of the order of some percent \cite{ferlaino_deform,vladimir}. Indeed, the combination of large magnetic moment and large mass in $^{167}$Er leads to a high value of its relative dipolar interaction strength. For example, under the circumstances such as particle numbers and trap frequencies as in Ref.~\cite{ferlaino_deform}, one would have $\epsilon_{\rm dd} = 0.15$ for $^{167}$Er, while for $^{53}$Cr $\epsilon_{\rm dd} = 0.02$ would hold. The experimental analysis presented in Ref.~\cite{ferlaino_deform}, however, is based on the possibility of carrying out the experiment for an arbitrary orientation of the dipoles, a feature which has not been considered in the present article for the low-lying excitation modes, as it represents a whole line of investigation by itself.

Further possibilities for studying systems with even stronger dipole-dipole interactions can be achieved via ultracold heteronuclear molecules such as $^{23}{\rm Na}^{40}{\rm K}$. These molecules could be cooled into their absolute rovibrational and hyperfine ground state by applying the stimulated Raman adiabatic passage (STIRAP) process \cite{bergmann}. They are particularly interesting for experimental studies of dipolar quantum gases due to some peculiar features. For example, they possess a large electric dipole moment of about $0.8$ Debye, leading to the substantial value of $\epsilon_{\rm dd} = 5.44$ for the dipolar interaction strength, provided the same particle numbers and trap frequencies from Ref.~\cite{ferlaino_deform} could be realized. Moreover, they are also chemically stable \cite{park}, allowing for relatively long lifetimes. Indeed, lack of chemical stability proved a prohibiting hurdle for KRb molecules, which were otherwise quite promising candidate systems for dipolar experiments \cite{K.-K.Ni10102008,arXiv:0811.4618,efficient_transfer}. At this point, is is important to mention the recent development of an experimental technique allowing for the manipulation of ultracold, rovibrational ground state NaK molecules \cite{silke_tune}. Thereby the strength, the sign, and the direction of the induced DDI becomes tunable, opening the road to experimentally probe a whole range of values of $\epsilon_{\rm dd}$.

The knowledge gained with a theoretical prediction of the real and imaginary parts of the low-lying excitation frequencies might also be used for an estimative of the number of oscillations that can be experimentally observed for a system with finite $\tau$. For atomic magnetic systems, for example, which are deeply in the CL regime due to the low values of $\epsilon_{\rm dd}$, this bears not much significance. For molecular systems, however, this can be useful. Indeed, even for magnetic dipolar interaction a molecular gas of erbium might lead to $\epsilon_{\rm dd} = 1.76$ \cite{er_molecule,vladimir}, which is considerably larger than the ones for the typical atomic magnetic systems discussed above.

We have considered $\epsilon_{\rm dd} = 1.33$, which can either be achieved by changing trap and atom number parameters or by tuning the interaction strength \cite{silke_tune}. Indeed, one oscillation takes place in $T=2\pi/{\rm Re}\Omega$ seconds. On the other hand, the oscillations decay with a time scale which is given by ${\mathcal T} = 1/{\rm Im} \Omega$. Thus, one can expect to observe about $n= {\rm Log}(4){\rm Re} \Omega/(2\pi {\rm Im} \Omega)$ oscillations before the amplitude is reduced, for example, to one fourth of its initial value. Combining results from Fig.~\ref{fig2} and from Fig.~\ref{fig3}, we can estimate that the monopole mode will take around four oscillations to decay by one fourth in its most damped regime, around $\overline{\omega}\tau \approx 0.2$, whereas more than twenty oscillations are due in a regime of finite relaxation time such as $\overline{\omega}\tau \approx 3.0$. For the other two modes, our result is that the most damped regime leads to a corresponding decay in just a few oscillations, and takes place at $\overline{\omega}\tau \approx 0.9$ for the radial quadrupole mode and at $\overline{\omega}\tau \approx 0.6$ for the three-dimensional quadrupole mode. For a regime of finite relaxation time such as $\overline{\omega}\tau \approx 3.0$, both modes are still strongly damped and only about a few, i.e. around five for the three-dimensinal quadrupole and three for the radial quadrupole mode, oscillations take place before the amplitudes decay to one fourth of its initial value. For this reason, we summarize by stressing that observing modes with a high oscillation frequency might be more promising than the ones with low damping rates.

\section{Conclusion} \label{CON}

We studied the low-lying excitations of a harmonically trapped three-dimensional Fermi gas featuring the long-range and 
anisotropic dipole-dipole interaction all the way from the collisionless to the hydrodynamic regime. Within the realm of the 
relaxation-time approximation, we were able to include the effects of collisions in the Boltzmann-Vlasov kinetic equation. 
In particular, we introduced 
the local equilibrium distribution, which corresponds to the hydrodynamic regime \cite{AristeuRapCom,Aristeu}, and we treated the relaxation 
time as a phenomenological parameter. Furthermore, we followed Ref.~\cite{String} and solved the BV equation by  
rescaling  appropriately both space and momentum variables of the equilibrium distribution. With this, we obtained ordinary differential equations of motion 
for the scaling parameters, whose linearization yields both the frequencies 
and the damping rates of the oscillations from the real and imaginary parts of complex frequencies, respectively.

In order to access the radial quadrupole mode in addition to the monopole and three-dimensional quadrupole mode, we 
started our calculation with a tri-axial configuration, which was later on specialized to the case of a cylinder-symmetric trap. 
The values of the frequencies that we found interpolate, as expected, between the values obtained previously in both the 
hydrodynamic regime \cite{AristeuRapCom,Aristeu} and the collisionless regime \cite{1367-2630-11-5-055017,PhysRevA.83.053628}, by increasing the 
relaxation time from zero to infinity.

By considering different values of the relaxation time, which could be achieved experimentally by means of different interaction 
strengths, for example, our analysis was able to identify both qualitative and quantitative features of the transition from the 
hydrodynamic to the collisionless regime. For particular values of the trap anisotropy and of the interaction strength, the 
transition might be smooth for one mode while being abrupt for another one. In view of the great precision with which measurements 
of the excitation frequencies in cold atomic systems are carried out nowadays, the present theoretical analysis could provide 
important information on the collisional properties of such systems.

A few questions remain open, which could be addressed with the help of the present theoretical framework. For example, the 
influence of the collisional term on the time-of-flight dynamics of the system could be considered, as this is a major 
diagnostic tool for cold atomic gases. To this end, however, it would be of major importance to determine microscopically the phenomenological introduced relaxation time \cite{ferlaino_out_eq}.
Moreover, the inclusion of finite-temperature effects on the analysis could also 
be of interest, as actual experiments are always performed at some finite, though very low temperature.

{\sl Note added.} Recently, an extensive study of the time-of-flight dynamics of a dipolar Fermi gas all the way from the collisionless to the hydrodynamic regime has been published, which is based on the theory presented here \cite{vladimir}.

\section*{Acknowlegements}
We acknowledge inspiring discussions with Antun Bala\v{z}, Vladimir Velji\'c, and Sandro Stringari. 
One of us (A. R. P. L.) acknowledges financial support from the Coordena\c{c}\~ao de Aperfei\c{c}oamento de Pessoal de N\'ivel Superior (CAPES) and from the Funda\c{c}\~ao Cearense de Apoio ao Desenvolvimento Cient\'ifico e Tecnol\'ogico (Grant No. BP2-0107-00129.02.00/15) as well as the hospitality of the Departments of Physics of the Federal University of Cear\'a (Brazil)
and of the Technische Universit\"at Kaiserslautern (Germany), where part of this work 
was carried out. Furthermore, we thank for the support from the German Research Foundation (DFG) via the Collaborative
Research Center SFB/TR49 {\it Condensed Matter Systems with Variable Many-Body Interactions} and, a final stage, via the Collaborative Research Center SFB/TR185 {\it Open System Control of Atomic and Photonic Matter}.
\begin{appendix}
\section{Computation of the Energy Integrals\label{computation}}
In order to make the paper self-contained, we present
in Appendix \ref{computation} the relevant steps for evaluating the Hartree-Fock energy integrals Eqs.~(\ref{Edf0}) and (\ref{Eex}) 
with the equilibrium distribution (\ref{ansatzwigner}). 
This leads to explicit expressions which only depend on the Thomas-Fermi radii and momenta 
$R_{\imath}$ and $K_{\imath}$, see also Refs.~\cite{miyakawa_t_2008,Aristeu}. A detailed derivation can also be found in Ref.~\cite{Falk_Diploma}.
\subsection{Hartree Energy}
Here, we sketch the calculation of the Hartree energy in the case of an equilibrium Wigner function of the form (\ref{ansatzwigner}). The basic idea behind the calculation of the Hartree integral (\ref{Edf0}) is to decouple the distribution functions and the 
interaction potential with respect to their spatial arguments via Fourier transforms. 
Thus, we rewrite the Hartree term by using the Fourier transform of the potential $\tilde{V}_{\rm d}({\bf k})$ according to
\begin{equation}
 E_{\rm d} =\frac{1}{2} \int \frac{d^3k''}{(2\pi)^3} \tilde{V}_{\rm d}({\bf k''}) \int \frac{d^3k}{(2\pi)^3} 
\tilde{\nu}^0({\bf -k''},{\bf k})  \int \frac{d^3k'}{(2\pi)^3} \tilde{\nu}^0({\bf k''},{\bf k'})\,.
\label{unterteiltEd}
\end{equation}
The first step is to compute the Fourier transform of the equilibrium Wigner function (\ref{ansatzwigner}). This is done by performing the integration in Cartesian coordinates over the six-dimensional sphere, which leads to
\begin{equation}
\tilde{\nu}^0(-{\bf k}'',{\bf k})=\frac{(2\pi)^{\frac{3}{2}}\overline{R}^3
h({\bf k})^{\frac{3}{4}}{\mathbf \Theta}[h({\bf k})]}{(k_x''^2R_x^2+k_y''^2R_y^2+k_z''^2R_z^2)^{\frac{3}{4}}}J_{\frac{3}{2}}\left[ 
h({\bf k})^{\frac{1}{2}}\left(k_x''^2R_x^2+k_y''^2R_y^2+k_z''^2R_z^2\right)^{\frac{1}{2}}  \right] \,, 
\label{FTW1}
\end{equation}
where $h({\bf k})=1-\sum_{\jmath} k_j^2/K_j^2$ is a suitable abbreviation and $J_{\imath}(x)$ is the $\imath$-th Bessel function of first kind. 

The next step is to integrate over the second variable in the Fourier-transformed Wigner function (\ref{FTW1}), which corresponds to the last two integrations in the three-fold three-dimensional integral in Eq.~(\ref{unterteiltEd}). To this end we perform a scaling transformation such that the problem becomes spherically symmetric and the angular part becomes trivial. The radial part can be dealt with by means of a trigonometric substitution and subsequent use of the identity \cite[(6.683)]{Grad}, leading to
\begin{equation}
\int \frac{d^3k}{(2\pi)^3}\tilde{\nu}^0({\bf k}'',{\bf k})=\frac{\overline{R}^3
\overline{K}^3}{(k_x''^2R_x^2+k_y''^2R_y^2+k_z''^2R_z^2)^{\frac{3}{2}}}  
J_3\left[ (k_x''^2R_x^2+k_y''^2R_y^2+k_z''^2R_z^2)^{\frac{1}{2}}  \right]\,. 
\label{FTD}
\end{equation}
Notice that the function at the right-hand side of Eq.~(\ref{FTD}) is even with respect to the components of ${\bf k}''$.

In order to be able to perform the last three-dimensional integration, one can proceed by means of a scaling which leads to a spherically symmetric argument of the Bessel function in Eq.~(\ref{FTD}). In this case, the radial part can be evaluated with he help of the identity \cite[(6.574.2)]{Grad}, while the angular part corresponds to the definition of the anisotropy function in Eq.~(\ref{definitionanistropyfunction}). Then, the dipolar interaction energy can be cast in the final form
\begin{equation}
E_{\rm d}=\frac{-48N^2 c_0}{8 \overline{R}^3}f\left( \frac{R_x}{R_z},\frac{R_y}{R_z} \right),
\end{equation}
with the constant (\ref{C0}).
\subsection{Fock Energy}
It is possible to compute the Fock integral (\ref{Eex}) along similar lines. Indeed, we start by rewriting the Fock term in the following form
\begin{equation}
E_{\rm ex}=-\frac{1}{2}\int d^3x' \int \frac{d^3k'}{(2\pi)^3}\int \frac{d^3k''}{(2\pi)^3} 
\overline{\tilde{\nu}}^0({\bf k''},{\bf x'}) \overline{\tilde{\nu}}^0(-{\bf k''},-{\bf x'})\tilde{V}_{\rm d}({\bf k'}) 
{\rm e}^{i{\bf x'}\cdot {\bf k'}} \, ,  
\label{generalfockequation}
\end{equation}
where $\overline{\tilde{\nu}}^0(-{\bf k}'',{\bf x}')$ denotes the Fourier transform of $\nu^0({\bf x},{\bf k})$ with respect to both variables
\begin{eqnarray}
\overline{\tilde{\nu}}^0(-{\bf k}'',{\bf x}')
=\int \frac{d^3k}{(2\pi)^3}{\rm e}^{i {\bf k}\cdot {\bf x}'}
\tilde{\nu}^0(-{\bf k}'',{\bf k})\, .
\end{eqnarray}
By using Cartesian coordinates, all three ${\bf k}$-integrals can be solved by an trigonometric substitution together by using the identity \cite[(6.688.2)]{Grad}. The final results then yields
\begin{equation}
\overline{\tilde{\nu}}^0(-{\bf k}'',{\bf x})=\frac{\overline{R}^3 \overline{K}^3}{\left[g({\bf k}'')
+z^2K_z^2+y^2K_y^2+x^2K_x^2\right]^{\frac{3}{2}}}J_3\left\{ [g({\bf k}'')+z^2K_z^2+y^2K_y^2+x^2K_x^2]^{\frac{1}{2}} \right\}\,,
\end{equation}
where $g({\bf k}'')=k_x''^2R_x^2+k_y''^2R_y^2+k_z''^2R_z^2$ is a suitable  abbreviation. 

It is clear that $\overline{\tilde{\nu}}^0({\bf k}'',{\bf x})$ is an even function with respect to ${\bf k}''$, which simplifies further calculations. The next step is to calculate the ${\bf x}'$-integral in Eq.~(\ref{generalfockequation}). 
In order to avoid a quadratic Bessel function, we use the integral representation \cite[(6.519.2.2)]{Grad}, thus leading to an integral over a Bessel function 
\begin{equation}
J_3^2\left\{\left[x^2K_x^2+y^2K_y^2+z^2K_z^2+g({\bf k}'')\right]^{\frac{1}{2}}\right\}=\frac{2}{\pi}\int_0^{\frac{\pi}{2}} dt 
J_6 \left\{ 2\sin t \left[ x^2K_x^2+y^2K_y^2+z^2K_z^2+g({\bf k}'') \right]^{\frac{1}{2}} \right\} \, .
\label{Besselidentity}
\end{equation}
Now, let us consider the three spatial integrals. They can all be solved with the help of a linear scaling together with the identity \cite[(6.726.2)]{Grad}. Thus, the solution of the ${\bf x}'$-integral reads
\begin{align}
\int d^3x' \overline{\tilde{\nu}}^0({\bf k}'',{\bf x'})^2 {\rm e}^{i{\bf k'}\cdot {\bf x'}}=\frac{2(2\pi)^{\frac{3}{2}}}{\pi}
\overline{R}^6\overline{K}^3\int_0^{\frac{\pi}{2}}\frac{dt}{(2\sin t)^6}\frac{\left( 4\sin^2 t -\frac{k_z'^2}{K_z^2}
-\frac{k_y'^2}{K_y^2}-\frac{k_x'^2}{K_x^2}\right)^{\frac{9}{4}}}{g({\bf k}'')^{\frac{9}{4}}} \nonumber \\
\times J_{\frac{9}{2}}\left[ g({\bf k}'')^{\frac{1}{2}}\left( 4\sin^2 t -\frac{k_z'^2}{K_z^2}-\frac{k_y'^2}{K_y^2}
-\frac{k_x'^2}{K_x^2}\right)^{\frac{1}{2}} \right] {\mathbf \Theta}\left(2\sin t-\sqrt{\frac{k_z'^2}{K_z^2}+\frac{k_y'^2}{K_y^2}
+\frac{k_x'^2}{K_x^2}}\,\right)\,.
\end{align}
The next step is to integrate the ${\bf k}''$-integral. Using the spherical symmetry the calculation of this integral can be done by substituting $u_i=k_i''R_{\imath}$ and by transforming afterwards these new integration variables into spherical coordinates.
This enables us to use the identity \cite[(6.561.17)]{Grad} and leads to
\begin{align}
\int d^3k''\int d^3x' \overline{\tilde{\nu}}^0({\bf k}'',{\bf x'})^2 {\rm e}^{i{\bf k'}\cdot {\bf x'}}=&\frac{\pi^2}{3}\int_0^{\frac{\pi}{2}}
\frac{dt}{(2 \sin t)^6}\left( 4\sin ^2t-\frac{k_z'^2}{K_z^2}-\frac{k_y'^2}{K_y^2}-\frac{k_x'^2}{K_x^2} \right)^3 \nonumber \\
&\times {\mathbf \Theta}\left( 2\sin t-\sqrt{\frac{k_z'^2}{K_z^2}+\frac{k_y'^2}{K_y^2}+\frac{k_x'^2}{K_x^2}} \,\right)\,. 
\label{doubleintdoublefourexp}
\end{align}
The last step of the calculation of the Fock term is to solve the ${\bf k}'$-integral. To this end we substitute 
$u_i=k_i'/K_{\imath}$ and switch to spherical coordinates. The integrals over the angular variables lead to the anisotropy function 
(\ref{definitionanistropyfunction}), and the radial and $t$-integrals can be solved in an elementary way, thus leading to the final result
\begin{equation}
E_{\rm ex}=\frac{48N^2c_0}{8\overline{R}^3}f\left( \frac{K_z}{K_x},\frac{K_z}{K_y} \right),
\end{equation}
for the Fock exchange contribution to the total energy of the system.
\section{Linearization \label{linearization}}

In Appendix \ref{linearization} we work out the linearization of the equations of motion (\ref{bequationcomplete}), (\ref{thetaequationcomplete}), (\ref{conn})--(\ref{couplinglocalequ}) for the respective scaling parameters $b_{\imath}, \Theta_{\imath}\textrm{ and }\Theta_{\imath}^{\rm le}$. To this end, we decompose all spatial and momentum scaling parameters around the equilibrium values $b_{\imath}^0=\Theta_{\imath}^0=\Theta_{\imath}^{{\rm le},0}=1$ for all $\imath$ according to Eq.~(\ref{linear1}). We start with summarizing the linearized equations for the local equilibrium Eqs.~(\ref{conn})--(\ref{couplinglocalequ}), which leads to 
\begin{eqnarray}
\delta \Theta_x^{\rm le}- \delta \Theta_y^{\rm le}&=&0\, ,\label{linequ1}\\
\sum_{\jmath} \delta b_{\jmath}+\delta \Theta^{\rm le}_x+\frac{1}{2}\delta \Theta_z^{\rm le}&=&0\, ,\\
A \sum_{\jmath} \delta b_{\jmath}-B\delta \Theta_x^{\rm le}+C \delta \Theta_z^{\rm le}&=&0\,, 
\label{linequ}
\end{eqnarray}
where $A$, $B$, and $C$ represent the following abbreviations 
\begin{eqnarray}
A&=&-\frac{48Nc_0}{2\prod_j R_{\jmath}} \left[ 2\frac{K_z}{K_x}f_1\left( \frac{K_z}{K_x},\frac{K_z}{K_y}\right)
+\frac{K_z}{K_y}f_2\left( \frac{K_z}{K_x},\frac{K_z}{K_y}\right) \right]\,,\\
B&=&\frac{\hbar^2K_x^2}{2M}+\frac{48Nc_0}{2\prod_j R_{\jmath}}\left[ \frac{K_z}{K_x}f_1\left( \frac{K_z}{K_x},\frac{K_z}{K_y}\right)
+ \frac{K_z^2}{K_x^2}f_{11}\left( 
\frac{K_z}{K_x},\frac{K_z}{K_y}\right)\nonumber \right. \\
&&
\left. +\frac{1}{2}\frac{K_z^2}{K_xK_y}f_{21}\left( \frac{K_z}{K_x},\frac{K_z}{K_y}\right)+\frac{K_z^2}{K_xK_y}f_{12}\left(\frac{K_z}{K_x},\frac{K_z}{K_y}\right)
+\frac{1}{2}\frac{K_z}{K_y}f_{2}\left(\frac{K_z}{K_x},\frac{K_z}{K_y}\right)
+\frac{1}{2}\frac{K_z^2}{K_y^2}f_{22}\left(\frac{K_z}{K_x},\frac{K_z}{K_y}\right)\right]\,,\\
C&=&\frac{\hbar^2 K_z^2}{2M}+\frac{48Nc_0}{2 \prod_j R_{\jmath}}\left[ \frac{K_z}{K_x}f_1\left( \frac{K_z}{K_x},\frac{K_z}{K_y}\right)
+ \frac{K_z^2}{K_x^2}f_{11}\left(
\frac{K_z}{K_x},\frac{K_z}{K_y}\right)+\frac{K_z^2}{K_xK_y}f_{12}\left( \frac{K_z}{K_x},\frac{K_z}{K_y}\right)
\right. \nonumber \\&&\left. 
+\frac{1}{2}\frac{K_z}{K_y}f_2\left( \frac{K_z}{K_x},\frac{K_z}{K_y}\right)
+\frac{1}{2}\frac{K_z^2}{K_xK_y}f_{21}\left( \frac{K_z}{K_x},\frac{K_z}{K_y}\right)
+\frac{1}{2}\frac{K_z^2}{K_y^2}f_{22}\left( \frac{K_z}{K_x},\frac{K_z}{K_y}\right)\right]\,.
\end{eqnarray}
Here, we have introduced the short-hand notations $f_1(x,y)=\partial f(x,y) / \partial x$, $f_2(x,y)=\partial f(x,y) / \partial y$ and 
$f_{\imath \jmath}$ stand for performing the $\imath$ and the $j$ derivative, respectively. From the linearized versions of the local equilibrium condition
(\ref{linequ1})--(\ref{linequ}) we obtain formulas, which reveal how the elongations for the scaling parameters 
$\delta \Theta_{\imath}^{\rm le}$ and $\delta b_{\imath}$ are related to each other
\begin{eqnarray}
\delta \Theta_x^{\rm le}&=&\frac{A-2C}{B+2C}\left(\sum_{\jmath} \delta b_{\jmath}\right)\,,
\label{thetaxle}\\
\delta \Theta_z^{\rm le}&=&-2\frac{A+B}{B+2C}\left(\sum_{\jmath} \delta b_{\jmath}\right)\,.
\label{thetayle}
\end{eqnarray}
Notice that, in the absence of the dipolar interaction, one would have $A=0$ and $B=C$, thus yielding $\delta \Theta_x^{\rm le}=\delta \Theta_z^{\rm le}$.

Inserting Eqs.~(\ref{thetaxle}) and (\ref{thetayle}) into the linearization of 
Eqs.~(\ref{bequationcomplete}) and (\ref{thetaequationcomplete}) the 6 parameters $\delta b_{\imath}$ 
and $\delta \Theta_{\imath}$ are determined by 6 algebraic equations. At first
we mention the analytical expressions for the elongations of the momentum scaling 
parameters $\delta \Theta_{\imath}$, which turn out to yield
\begin{equation}
\delta \Theta_x= \delta \Theta_y \label{iso_mom_delta}
\end{equation}
due to the cylinder symmetry in momentum space, with the equation for $\delta \Theta_x$ being given by
\begin{equation}
\delta \dot{\Theta}_x+2 \delta \dot{b}_x=-\frac{1}{\tau}\left\{\delta \Theta_x-\left[\frac{A-2C}{B+2C}\left(\sum_{\jmath} \delta b_{\jmath}\right)\right]\right\}\,, \label{deltadotthetax}
\end{equation}
wheareas the equation for $\delta \Theta_z$ reads
\begin{equation}
\delta \dot{\Theta}_z+2 \delta \dot{b}_z=-\frac{1}{\tau}\left[\delta \Theta_z+2\frac{A+B}{B+2C}\left(\sum_{\jmath} \delta b_{\jmath}\right)\right]\,. \label{deltadotthetay}
\end{equation}
The equations for the elongations of the spatial scaling parameters $\delta b_{\imath}$ finally read
\begin{equation}
  \delta \ddot{b}_{\imath}+\sum_{\jmath} O_{\imath \jmath}\delta b_{\jmath}+\sum_{\jmath} D_{\imath \jmath}\delta \Theta_{\jmath}=0\,, 
\label{deltadotdotb}
\end{equation}
where we have defined
\begin{eqnarray}
O_{xx}&=&\omega_x^2+\frac{\hbar^2 K_x^2}{M^2 R_x^2}-\frac{48Nc_0}{MR_x^2 \prod_j R_{\jmath}} E_1, \quad
O_{xy}=-\frac{48Nc_0}{MR_x^2\prod_j R_{\jmath}}F_{12}, \quad
O_{xz}=-\frac{48Nc_0}{MR_x^2\prod_j R_{\jmath}} G_{12} , \nonumber\\
O_{yx}&=&-\frac{48Nc_0}{MR_y^2\prod_jR_{\jmath}} F_{21}, \quad
O_{yy}=\omega_y^2+\frac{\hbar^2K_y^2}{M^2R_y^2}
-\frac{48Nc_0}{MR_y^2\prod_j R_{\jmath}}E_2, \quad
O_{yz}=-\frac{48Nc_0}{MR_y^2 \prod_j R_{\jmath}} G_{21}, \nonumber\\
O_{zx}&=&-\frac{48Nc_0}{MR_z^2\prod_j R_{\jmath}}I_{12}, \quad
O_{zy}=-\frac{48Nc_0}{MR_z^2\prod_j R_{\jmath}}I_{21}, \quad
O_{zz}= \omega_z^2+\frac{\hbar^2 K_z^2}{M^2 R_z^2}
-\frac{48Nc_0}{MR_z^2 \prod_j R_{\jmath}}J, \nonumber\\
D_{xx}&=&- \frac{\hbar^2K_x^2}{M^2 R_x^2}-\frac{48Nc_0}{MR_x^2 \prod_j R_{\jmath}} \frac{1}{2} \frac{K_z^2}{K_x^2} f_{11}
\left( \frac{K_z}{K_x},\frac{K_z}{K_y}\right), \quad
D_{xy}=-\frac{48Nc_0}{MR_x^2 \prod_j R_{\jmath}} H_{12}, \nonumber\\
D_{xz}&=&\frac{48Nc_0}{MR_x^2 \prod_j R_{\jmath}}\left[\frac{1}{2}\frac{K_z^2}{K_x^2}f_{11}\left( \frac{K_z}{K_x},\frac{K_z}{K_y}\right)+H_{12}\right], \quad
D_{yx}=-\frac{48Nc_0}{MR_y^2\prod_j R_{\jmath}} H_{21}, \nonumber\\
D_{yy}&=&-\frac{\hbar^2 K_y^2}{M^2R_y^2}-\frac{48Nc_0}{MR_y^2\prod_j R_{\jmath}}
\frac{1}{2} \frac{K_z^2}{K_y^2}f_{22}\left( 
\frac{K_z}{K_x},\frac{K_z}{K_y}\right) , \quad
D_{yz}=\frac{48Nc_0}{MR_y^2\prod_j R_{\jmath}}\left[\frac{1}{2}\frac{K_z^2}{K_y^2} f_{22}\left( \frac{K_z}{K_x},\frac{K_z}{K_y}\right)+H_{21}\right] , \nonumber\\
D_{zx}&=&\frac{48Nc_0}{MR_z^2 \prod_j R_{\jmath}}M_{12}, \quad
D_{zy}=\frac{48Nc_0}{MR_z^2 \prod_j R_{\jmath}}M_{21}, \quad
D_{zz}=-\frac{\hbar^2K_z^2}{M^2R_z^2}-\frac{48Nc_0}{MR_z^2 
\prod_j R_{\jmath}}\left( M_{12}+M_{21} \right)\,.
\end{eqnarray}
Here, we have introduced the following abbreviations
\begin{eqnarray}
E_{\imath}&=&2f\left( \frac{R_x}{R_z},\frac{R_y}{R_z}\right)-2\frac{R_{\imath}}{R_z}f_{\imath}
\left( \frac{R_x}{R_z},\frac{R_y}{R_z}\right)-2f\left( \frac{K_z}{K_x},\frac{K_z}{K_y}\right) 
+2\frac{K_z}{K_{\imath}}f_{\imath}\left( \frac{K_z}{K_x},\frac{K_z}{K_y}\right)+\frac{R_{\imath}^2}{R_z^2}f_{\imath \imath}\left( \frac{R_x}{R_z},\frac{R_y}{R_z}\right)\,,\nonumber\\
F_{\imath \jmath}&=&f\left( \frac{R_x}{R_z},\frac{R_y}{R_z}\right)-\frac{R_{\imath}}{R_z}f_{\imath}
\left( \frac{R_x}{R_z},\frac{R_y}{R_z}\right)-f\left( \frac{K_z}{K_x},\frac{K_z}{K_y}\right)
+\frac{K_z}{K_{\imath}}f_{\imath}\left( \frac{K_z}{K_x},\frac{K_z}{K_y}\right)-\frac{R_{\jmath}}{R_z}f_{\jmath}\left( \frac{R_x}{R_z},\frac{R_y}{R_z}\right)\nonumber \nonumber\\
&&+\frac{R_{\imath}R_{\jmath}}{R_z^2}f_{\imath \jmath}\left( \frac{R_x}{R_z},\frac{R_y}{R_z}\right),\nonumber\\
G_{\imath \jmath}&=&f\left( \frac{R_x}{R_z},\frac{R_y}{R_z}\right)-\frac{R_{\imath}}{R_z}f_{\imath}
\left( \frac{R_x}{R_z},\frac{R_y}{R_z}\right)-f\left( \frac{K_z}{K_x},\frac{K_z} {K_y}\right)
+\frac{K_z}{K_x}f_{\imath}\left( \frac{K_z}{K_x},\frac{K_z}{K_y}\right)+\frac{R_{\jmath}}{R_z}f_{\jmath}\left( \frac{R_x}{R_z},\frac{R_y}{R_z}\right) \nonumber \nonumber\\
& &-\frac{R_{\imath}^2}{R_z^2}f_{\imath \imath}\left( \frac{R_x}{R_z},\frac{R_y}{R_z}\right)-\frac{R_{\imath}R_{\jmath}}{R_z^2}f_{\imath \jmath}\left( \frac{R_x}{R_z},\frac{R_y}{R_z}\right)\, ,\nonumber\\
H_{\imath \jmath}&=&-\frac{1}{2}\frac{K_z}{K_j}f_{\jmath}\left( \frac{K_z}{K_x},\frac{K_z}{K_y}\right)
+\frac{1}{2}\frac{K_z^2}{K_{\imath}K_j}f_{\imath \jmath}\left( \frac{K_z}{K_x},\frac{K_z}{K_y}\right)\,,\nonumber\\
I_{\imath \jmath}&=&f\left( \frac{R_x}{R_z},\frac{R_y}{R_z}\right)+\frac{R_{\jmath}}{R_z}f_{\jmath}
\left( \frac{R_x}{R_z},\frac{R_y}{R_z}\right)-f\left( \frac{K_z}{K_x},\frac{K_z}{K_y}\right)
-\frac{K_z}{K_{\imath}}f_{\imath}\left( \frac{K_z}{K_x},\frac{K_z}{K_y}\right) \nonumber \\
&&-\frac{K_z}{K_j} f_{\jmath}\left( \frac{K_z}{K_x},\frac{K_z}{K_y}\right)
-\frac{R_{\imath}}{R_z}f_{\imath}\left( \frac{R_x}{R_z},\frac{R_y}{R_z}\right)-\frac{R_{\imath}^2}{R_z^2}f_{\imath \imath}\left( 
\frac{R_x}{R_z},\frac{R_y}{R_z}\right)-\frac{R_{\jmath}R_{\imath}}{R_z^2}f_{\jmath \imath}\left( \frac{R_x}{R_z},\frac{R_y}{R_z}\right)\,,\nonumber\\
J&=&2f\left( \frac{R_x}{R_z},\frac{R_y}{R_z}\right)+4\frac{R_x}{R_z} 
f_1\left( \frac{R_x}{R_z},\frac{R_y}{R_z}\right)+4\frac{R_y}{R_z}f_2\left( \frac{R_x}{R_z},
\frac{R_y}{R_z}\right) \nonumber \\
& &-2f\left( \frac{K_z}{K_x},\frac{K_z}{K_y}\right)-2\frac{K_z}{K_x}f_1\left( 
\frac{K_z}{K_x},\frac{K_z}{K_y}\right)-2\frac{K_z}{K_y}f_2\left( \frac{K_z}{K_x},
\frac{K_z}{K_y}\right)+\frac{R_x^2}{R_z^2}f_{11}\left( \frac{R_x}{R_z},\frac{R_y}{R_z}\right) \nonumber \\
& &+\frac{R_xR_y}{R_z^2}f_{12}\left( \frac{R_x}{R_z},\frac{R_y}{R_z}\right)+\frac{R_xR_y}{R_z^2}f_{21}
\left( \frac{R_x}{R_z},\frac{R_y}{R_z}\right)+\frac{R_y^2}{R_z^2}
f_{22}\left( \frac{R_x}{R_z},\frac{R_y}{R_z}\right)\,,\nonumber\\
M_{\imath \jmath}&=&\frac{K_z}{K_{\imath}}f_{\imath}\left( \frac{K_z}{K_x},\frac{K_z}{K_y}\right)+\frac{1}{2} 
\frac{K_z^2}{K_{\imath}^2}f_{\imath \imath}\left( \frac{K_z}{K_x},\frac{K_z}{K_y}\right)+\frac{1}{2}
\frac{K_z^2}{K_{\imath}K_j}f_{\jmath \imath}\left( \frac{K_z}{K_x},\frac{K_z}{K_y}\right)\, ,
\end{eqnarray}
where $R_1=R_x,R_2=R_y,K_1=K_x$, $K_2=K_y$ and $i,j \in \{ 1,2\}$.

Thus, we conclude that the elongations $\delta b_{\imath}$, $\delta \Theta_{\imath}$, and $\delta \Theta_{\imath}^{\rm le}$ out of equilibrium are determined by (\ref{linequ1}) and (\ref{thetaxle})--(\ref{deltadotdotb}).
\end{appendix}

\end{document}